\documentclass[preprint,authoryear]{elsarticle}

\usepackage[section]{placeins}
\usepackage{amsmath}
\usepackage{todonotes}

\usepackage{amssymb}

\usepackage{lineno}

\journal{Ocean Modelling}

\begin{document}
\begin{frontmatter}



\title{
Accuracy and stability analysis of horizontal discretizations used in unstructured grid ocean models
}

\author[IME,MP]{Fabricio Rodrigues Lapolli}
\author[IME]{Pedro da Silva Peixoto}
\author[MP]{Peter Korn}

\affiliation[IME]{organization={Universidade de Sao Paulo},
            addressline={Rua do Matao, 1010, Cidade Universitaria}, 
            city={Sao Paulo},
            postcode={05508-090}, 
            state={Sao Paulo},
            country={Brazil}}

\affiliation[MP]{organization={Max-Planck Institute for Meteorology},
            addressline={Bundestrasse, 53}, 
            city={Hamburg},
            postcode={20146}, 
            state={},
            country={Germany}}

\begin{abstract}
    One important tool at our disposal to evaluate the robustness of Global Circulation Models (GCMs) is to understand the horizontal discretization of the dynamical core under a shallow water approximation. Here, we evaluate the accuracy and stability of different methods used in, or adequate for, unstructured ocean models considering shallow water models. Our results show that the schemes have different accuracy capabilities, with the A- (NICAM) and B-grid (FeSOM 2.0) schemes providing at least 1st order accuracy in most operators and time integrated variables, while the two C-grid (ICON and MPAS) schemes display more difficulty in adequately approximating the horizontal dynamics. Moreover, the theory of the inertia-gravity wave representation on regular grids can be extended for our unstructured based schemes, where from least to most accurate we have: A-, B, and C-grid, respectively. Considering only C-grid schemes, the MPAS scheme has shown a more accurate representation of inertia-gravity waves than ICON. In terms of stability, we see that both A- and C-grid MPAS scheme display the best stability properties, but the A-grid scheme relies on artificial diffusion, while the C-grid scheme doesn't. Alongside, the B-grid and C-grid ICON schemes are within the least stable. Finally, in an effort to understand the effects of potential instabilities in ICON, we note that the full 3D model without a filtering term does not destabilize as it is integrated in time. However, spurious oscillations are responsible for decreasing the kinetic energy of the oceanic currents. Furthermore, an additional decrease of the currents' turbulent kinetic energy is also observed, creating a spurious mixing, which also plays a role in the strength decrease of these oceanic currents.
\end{abstract}










\end{frontmatter}


\section{Introduction}
\label{sec:Introduction}

Much of the scientific knowledge of the climate is largely due to the development of Earth System Models (ESMs), i.e. coupled models consisting of the atmosphere, ocean, sea ice, and land surface. The ocean, in particular, is a key component of these ESMs and a driver of the climate. Consequently, it is crucial to develop and improve such ocean models, with particular attention to global models \citep{Randalletal2019,FoxKemperetal2019}.

These efforts, along with the atmospheric modelling community, allowed us to acquire important insights related to these numerical models, such as being able to compartmentalize models into what is termed dynamical cores along with several physical parametrizations \citep{Thuburn2008, StaniforthThuburn2012}. Combined, these form the main building blocks of the current operational ESMs. The dynamical core is defined as being responsible for solving the governing equations on the resolved scales of our domain \citep{Randalletal2019, Thuburn2008}. For climate modelling, it is important that these cores are able to mimic important physical properties, such as mass and energy conservation, minimal grid imprinting, increased accuracy, and reliable representation of balanced and adjustment flow, which can be achieved by using a proper grid geometry and horizontal discretization \citep{StaniforthThuburn2012}. However, the use of unstructured grids may pose challenges in fulfilling these properties. 



Traditional ocean models commonly used Finite Difference or Finite Volume discretization on regular structured grids \citep{Fox-Kemperetal2019}, e.g. NEMO \citep{Madecetal2022}, MOM6 \citep{Adcroftetal2019}. This approach was useful for the limited regional modelling. However, for global models it posed some problems. The most critical is the presence of singularity points at the poles, which constrained the timestep size for explicit methods, potentially making it unfeasible for use in high resolution models \citep{Sadourny1972, StaniforthThuburn2012, Randalletal2019}. Therefore,  in recent years, a lot of effort has been put on the development of unstructured global oceanic models.


Given the success of triangular grids on coastal ocean models, one popular approach is the use of triangular icosahedral-based global models, i.e. using geodesic triangular grids. However, there are still present issues with triangular grids, in particular with the variable positioning considering a C-grid staggering. The C-grid staggering \citep{ArakawaLamb1977} considers the velocity decomposed into normal components at the edges of a computational cell. On traditional quadrilateral meshes, this staggering was found to more accurately represent the inertia-gravity waves \citep{Randall1994}. On unstructured triangular grids, a spurious oscillation is present on the divergence field manifested as a \textit{chequerboard pattern}, and it is present due to the excessive degrees of freedom (DOF) on the vector velocity field \citep{Gassmann2011, LeRouxetal2005, Danilov2010, Welleretal2012}. In theory, these can lead to incorrect results if not correctly filtered, or can potentially trigger instabilities.

This \textit{chequerboard pattern} issue led modellers to avoid triangular grids. One potential solution, which is used by MPAS-O model, is to use the dual grid, based on hexagonal-pentagonal cells, formed by connecting the circumcentres of the triangles (defining a Voronoi grid dual to the triangulation). By relying on the orthogonality properties between the triangular and the dual quasi-hexagonal grid, the problem of the spurious divergence modes is avoided. However, the noise will appear on the vorticity field, where it is easier to filter \citep{Welleretal2012}. 

Another potential solution to the chequerboard pattern on triangular grids is the use of filters on the divergence field in order to dampen these oscillations. However, these can potentially break the conservative properties of the model. A solution devised by the ICON-O ocean model community is the implementation of mimetic operators that required the preservation of some physical dynamical core properties, while, simultaneously, filtering the noise of the divergence field  \citep{KornDanilov2017, Korn2017, KornLinardakis2018}. However, the added triangle distortion of the grid might not completely remove the noise, and, thus, the filtering property might be at most approximate.

In order to avoid the noise on the divergence field of triangular grids at all, a possibility is to avoid C-grid staggering. FeSOM 2.0 model, for example, uses the (quasi-) B-grid discretization in which the velocity vector field and the height field are allocated at the cells centre and vertices, respectively \citep{Danilovetal2017}. Alternatively, the NICAM atmospheric model, uses the A-grid discretization, which has all its fields positioned at the vertices of the grid \citep{Tomitaetal2001, TomitaSatoh2004}. Nonetheless, there are drawbacks from this solution. For example,  both staggerings display spurious modes that are potentially unstable without treatment \citep{Randall1994}. The nature of these modes differs for each of the grid designs. The A-grid source of numerical noise is related to the manifestation of spurious pressure modes, whilst the B-grid allows the manifestation of spurious inertial modes due to excessive DOFs of the horizontal velocity \citep{Tomitaetal2001, Danilovetal2017}.

Nonetheless, regardless of grid design, other artefacts may also be present. One particular spurious oscillation was detected on an energy-enstrophy conserving scheme (EEN) on an atmospheric model, leading to an instability \citep{Hollingsworth1983}. This kind of instability is dependent on the fastest internal modes of the model,  the horizontal velocity and resolution of the model \citep{Belletal2017}. Due to the presence of distortion on these newer models, instability might be more easily triggered \citep{Peixotoetal2018}. This kind of noise is noticeable on atmospheric models, due to the large flow speeds of the atmosphere and the near  to kilometre grid resolutions used in their simulations \citep{Skamarocketal2012}. Although the ocean dynamics are less energetic than the atmosphere, the higher distortion of the grids and the rapid increase of resolution towards submesoscale models make the effects of this noise more relevant. In fact, some models, such as the NEMO's EEN ocean model, identified this noise and its effects, which have shown significant effects on the model's mesoscale jets and submesoscale phenomena \citep{Ducoussoetal2017}.

Considering the challenges discussed, this works aims at investigating and comparing the accuracy and stability of different horizontal discretizations used in global unstructured ocean models. First, in contrast to regular grids, the unstructured nature of the mesh may play a role in the computation of the underlying operators of each scheme's staggering design. Similarly, regular grids have a well-known inertia-gravity wave dispersion, therefore, can we expect a similar behaviour for the schemes in unstructured grids. Finally, these unstructured grid schemes are prone to instabilities due to their discretization, therefore, their different designs might play a role in their overall stability.

To address these questions, we chose to evaluate both MPAS-O and ICON-O C-grid discretization schemes, due to their robustness and different approaches on computing the necessary operators; the FESOM2.0 for the B-grid scheme; and the NICAM A-grid scheme, which, to our knowledge, currently is not present in ocean models, but could be easily incorporated in existing ones. The investigation will be mostly focused on the rotating shallow water system of equations, but we will also evaluate some properties of the 3D ICON-O model. In section \ref{sec:SWM}, we describe each of the aforementioned schemes. In section \ref{sec:IdeaTestCase}, we evaluate the accuracy and rate of convergence of each of these schemes. In section \ref{subsec:SWTI}, we perform a time integration, in order to evaluate the accuracy of the integrated quantities and to observe some important properties of the models, such as the representation of inertia-gravity waves and the manifestation of near-grid scale oscillations under near realistic conditions. Finally, we evaluate the stability of the models under the effects of spurious grid scale oscillations and the effects of these oscillations in a 3D realistic oceanic ICON-O model.

\FloatBarrier
\section{Shallow Water models}
\label{sec:SWM}

In order to investigate these models, we test the schemes under the shallow water system of equations \citep{Gill1982}. This system is as follows:

\begin{subequations}
    \label{eq:NLSWE}
    \begin{align}
        \frac{\partial h}{\partial t} &= -\nabla\cdot(\mathbf{u}h)  \\
        \begin{split}
            \frac{\partial \mathbf{u}}{\partial t} &=  -\mathbf{u}\cdot\nabla\mathbf{u}-\nabla\Phi -f \mathbf{u}^\perp + F \\
            &= -\nabla(\Phi + E_k) -\omega u^\perp + F
        \end{split} \label{subeq:SWvel}
    \end{align}
\end{subequations}
where $h$ and $\mathbf{u}$ are the height (scalar) and velocity (vector) fields of the system; $f$ is the Coriolis parameter; $\omega=\zeta+f$ is the absolute vorticity; $\zeta$ is the relative vorticity or curl; $\Phi=g(b+h)$ is the geopotential, $g$ is the acceleration of gravity, and $b$ is the bathymetry; $\mathbf{u}^{\perp} = \mathbf{\hat{k}}\times\mathbf{u}$ is the perpendicular vector field $\mathbf{u}$ and $\mathbf{\hat{k}}$ is the vertical unit vector; and $E_k=\vert\mathbf{u}\vert^2/2$ is the kinetic energy. The right-hand most side of \eqref{subeq:SWvel} is known as the vector invariant form of the system of equations.

On this section, we present an introduction to each model and how they interpolate their quantities of the shallow water operators. On the next section, Section \ref{sec:IdeaTestCase}, we describe how each model compute each of the shallow water operator.

\subsection{Discrete Framework}

\begin{figure}[h]
    \centering
    \includegraphics[width=.6\linewidth]{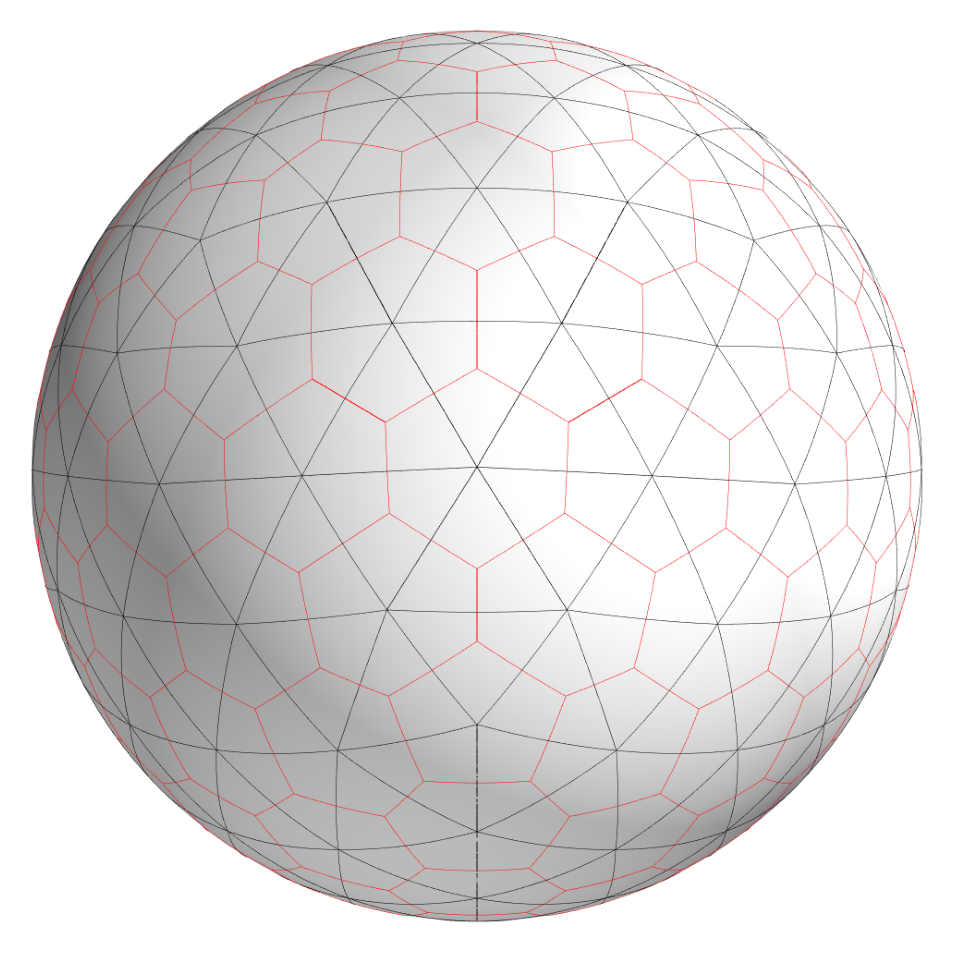}
    \caption{SCVT primal (black lines) and dual (red lines) g$_2$ grid.}
    \label{fig:gridg2}
\end{figure}

The models were evaluated with the Spherical Centroidal Voronoi Tessellation (SCVT) optimization \citep{MiuraKimoto2005} between the second (g$_2$) and eighth (g$_8$) refinements of the icosahedral grid (Table \ref{tb:SCVTspatres}). This optimization has the property of having its vertices coincide with the barycentre of the dual cells, quasi-hexagonal (red lines of Figure \ref{fig:gridg2}). This allows for an increase of accuracy for operators defined on vertices. This choice was made for simplicity, but may unfairly benefit both NICAM and MPAS-O model. However, ICON-O typically favours Spring Dynamics Optimization \citep{Kornetal2022}, which increase the convergence of some grid properties, such as reduction of mesh distortion, convergence of edge midpoints \citep{MiuraKimoto2005}.

\begin{table}[h]
    \centering
    \begin{tabular}{c|c c}
        \hline
         & Circ. distance (Km) & Edge length (Km)\\
        \hline
         g$_2$& 1115 & 1913\\
         g$_3$& 556  & 960\\
         g$_4$& 278  & 480\\
         g$_5$& 139  & 250\\
         g$_6$& 69   & 120\\
         g$_7$& 35   & 60\\
         g$_8$& 17   & 30\\
        \hline
    \end{tabular}
    \caption{Spatial resolution of the SCVT grid, considering the average distance between triangles circumcentre and the average edge length in Km.}
    \label{tb:SCVTspatres}
\end{table}

The structure of the grid domain will consist of triangular cells (primal grid)  $K\in \mathcal{C}$ with edges $e\in \mathcal{E}$. The set of edges of a particular cell $K$ is represented by $\partial K$. The vertices in the endpoint of these edges are represented by $\partial e$. Occasionally, when necessary, the edges may be denoted as $e=K\vert L$ where it is positioned between cells $K$ and $L$. The dual cells will be denoted by the $\widehat{(\cdot)}$ symbol. The dual cells and edges, for example, are denoted as $\widehat{K} \in \widehat{\mathcal{C}}$ and $\hat{e}\in \widehat{\mathcal{E}}$, respectively. Furthermore, the centre/midpoint position of the elements will be denoted by the boldface, e.g. the cell circumcentre position $\mathbf{K}$, and the length or area of the respective elements will be denoted by $\vert\cdot\vert$, e.g. $\vert e\vert, \ \vert\widehat{K}\vert$ is the edge length and dual cell area, respectively.

We note that the relationship between the primal and dual mesh will differ depending on the model discretization definitions. Some models use circumcentre of the triangle to construct the dual mesh. The resulting relationship will be a Delaunay triangulation (for the primal) and a Voronoi diagram (for the dual), making their edges orthogonal to each other, which can be exploited by these models.

Additionally, normal ($\mathbf{n}_e$) and tangent ($\mathbf{t}_e$) vectors are positioned at the edge $\mathbf{e}$ or $\hat{\mathbf{e}}$, such that $\mathbf{n_e}\times\mathbf{t_e} = \mathbf{e}$. The former vector is normal to $e$, while the latter is parallel to it. These definitions are summarized in Table \ref{tab:griddef}.

\begin{table}[h]
    \centering
    \begin{tabular}{c c}
        \hline
        Symbol & Description \\
        \hline
         $\mathcal{C}$ & Set of primal cells \\
         $\mathcal{E}$ & Set of primal edges \\
         $K$, $L$ &  primal grid cells \\
         $\partial K$ &  Set of edges of cell $K$ \\
         $e = K\vert L$ & primal edge \\
         $n_e$ , $t_e$ &  Normal and tangent vectors on edge $e$ \\
         $\partial e$ &  Set of vertices of edge $e$ \\
         \hline
         $\widehat{\mathcal{C}}$ & Set of dual cells \\
         $\widehat{\mathcal{E}}$ & Set of dual edges \\
         $\widehat{K}$, $\widehat{L}$ &  dual grid cells \\
         $\partial\widehat{K}$ &  Set of edges of cell $\widehat{K}$ \\
         $\hat{e} = \widehat{K}\vert\widehat{L}$ & dual edge \\
         $n_{\hat{e}}$ , $t_{\hat{e}}$ &  Normal and tangent vectors on edge $\hat{e}$ \\
         $\partial\hat{e}$ &  Set of vertices of edge $\hat{e}$ \\
        \hline
    \end{tabular}
    \caption{Definitions of the grid structure.}
    \label{tab:griddef}
\end{table}

\subsection{NICAM (A-grid)}

The NICAM model is a non-hydrostatic atmospheric-only model developed at AICS, RIKEN. Its development aimed to develop a high-performance global model \citep{TomitaSatoh2004}. The model has been shown to produce accurate results for simulations with a 3.5 km mesh size, and recent developments aim to pursue sub-kilometre grid scales \citep{Miyamotoetal2013}.

\begin{figure}[h]
    \centering
    \includegraphics[width=.5\linewidth]{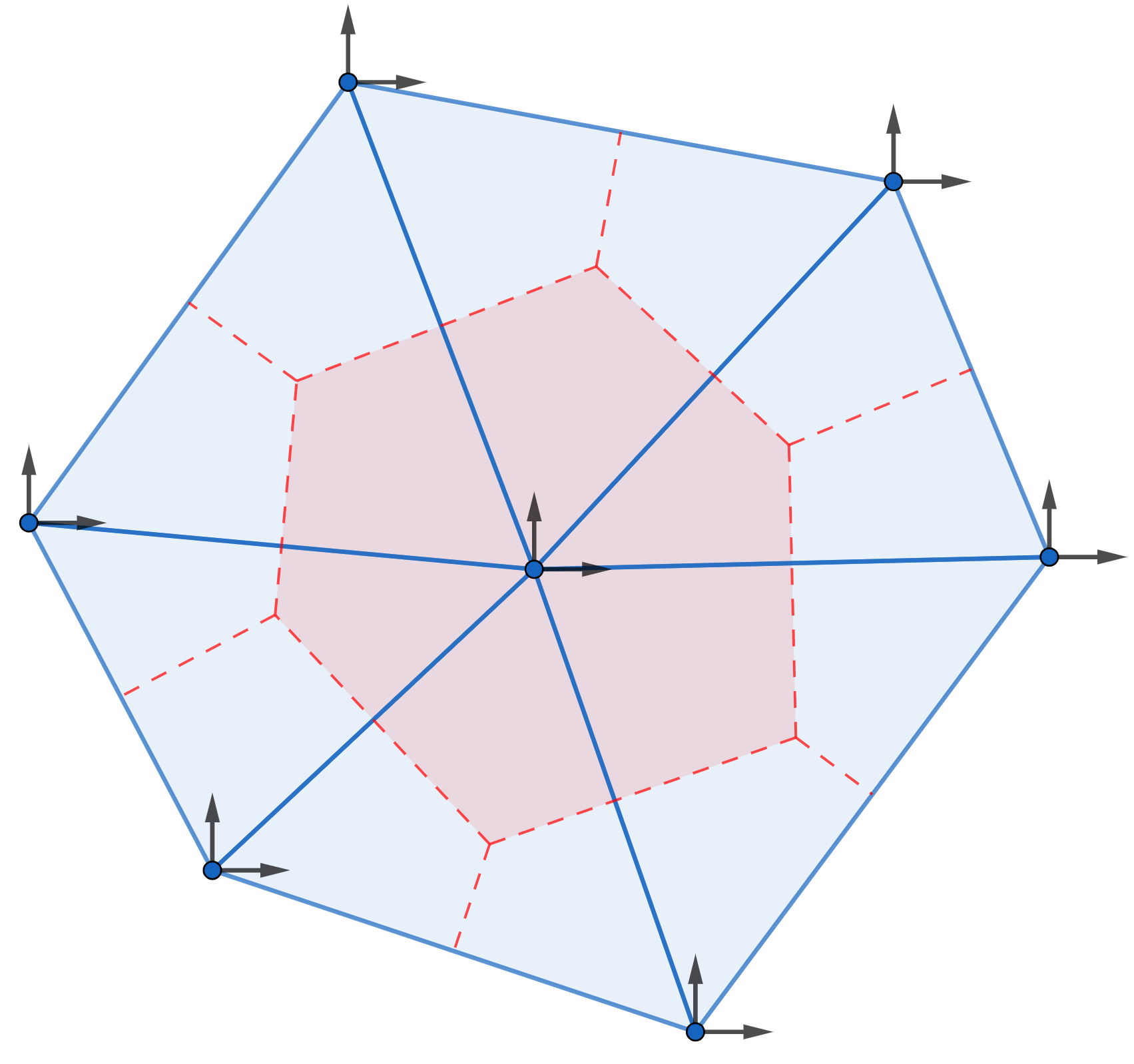}
    \caption{A-grid cell structure. The blue circles on the vertices are the height scalars points and the arrows are the components of the velocity vector points.}
    \label{fig:Agrid}
\end{figure}

NICAM's dynamical core's horizontal component is based on the A-grid discretization, in which all variables are located at the grid vertices (Figure \ref{fig:Agrid}). The discretization of this scheme allows only for mass conservation. Other quantities, specially related to the velocity equation, can not be conserved. This is because this scheme allows for spurious pressure modes, which may destabilize the model, thus, requiring filtering.

Additionally, small scale oscillations may also be present due to the grid imprinting, which may also decrease the model's stability \citep{Tomitaetal2001}. These oscillations, however, can be remedied with a proper grid optimization. One important requirement is that the dual cell centre coincide centre of mass coincide with the vertex of the grid, guaranteeing consistency of the discretization of the operators.

Moreover, NICAM's A-grid discretization compared to the MPAS-O shallow water scheme this scheme has been shown to display a higher resilience when non-linearities are present, implying that it can better treat some types of instabilities than other models \citep{Yuetal2020}. Therefore, despite this scheme not have originally been developed for oceanic purposes, It can be suitably implemented in such applications.

\subsubsection{Interpolating operators}

To compute the operations in the shallow water system, we need that the position of these operators coincide with the variables, i.e., at the vertices. Therefore, the computation must be performed on the dual cell. To do this, it is necessary to interpolate the variables at the dual edge midpoint. We do this by first interpolating at the circumcentre of the primal cell:

\begin{subequations}
    \label{eq:AScavtoK}
    \begin{equation}
        \widetilde{h}^K = \frac{1}{\vert K \vert}\sum_{v\in \partial e_K} w_v h_v,
    \end{equation}
    \begin{equation}
        \widetilde{\mathbf{u}}^K = \frac{1}{\vert K \vert}\sum_{v\in e_K} w_v \mathbf{u}_v,
    \end{equation}
\end{subequations}
where $w_v$ is the sectional triangular area formed by the circumcentre and the opposite vertices of the cell (See Figure 2 of \cite{Tomitaetal2001}). This interpolation, known as the barycentric interpolation, will provide us with a second order accurate interpolation. A second order interpolation to the edge midpoint can then be met by averaging neighbouring primal cells:
\begin{subequations}
    \label{eq:AScaKtoe}
    \begin{equation}
        \widetilde{h}^{\hat{e}} = \frac{1}{2}(h_K + h_L), \\
    \end{equation}
    \begin{equation}
        \widetilde{\mathbf{u}}^{\hat{e}} = \frac{1}{2}(\mathbf{u}_K+\mathbf{u}_L).
    \end{equation}
\end{subequations}

\subsection{FESOM (B-grid)}

FESOM 2.0, developed in the Alfred Wegener Institute, contains ocean \citep{Danilovetal2017} and ice \citep{Danilovetal2015, Danilovetal2023} components only. The model is an update from its previous 1.4 model \citep{Wangetal2008}. The new model was developed to provide faster simulations compared to its 1.4 predecessor \citep{Scholzetal2019}, which is partly owed to the change from Finite Element Methods to Finite Volume discretization \citep{Danilovetal2017}.

In addition to its updated components and faster simulations, FESOM 2.0's horizontal discretization of the dynamical core is based on the Arakawa B-grid staggering \citep{ArakawaLamb1977}. It is important to note that there is no true analogue of the B-grid on triangles \citep{Danilov2013}, and such a discretization has been coined as quasi-B-grid. However, due to the similarities in the positioning of the fields in the cell, in this work, we will describe this discretization only as B-grid.

Contrary to the aforementioned A-grid, this discretization is free of pressure modes. However, it allows for the presence of spurious inertial modes, due to its excessive degrees of freedom \citep{Danilovetal2017}. Thus, again, requiring the use of filters to remove these oscillations.

\begin{figure}[h]
    \centering
    \includegraphics[width=.5\linewidth]{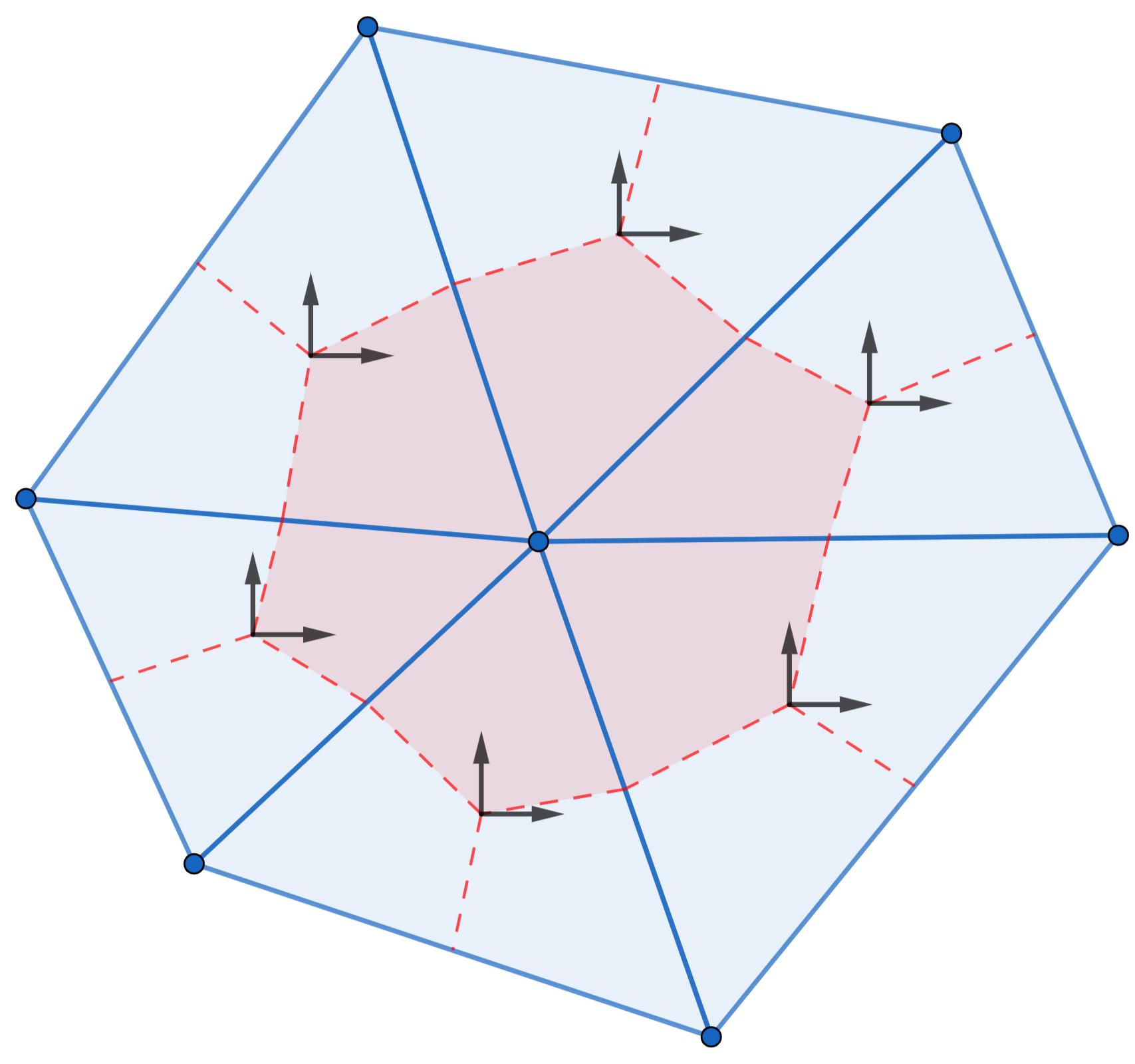}
    \caption{B-grid cell structure. The blue circles on the vertices are the height scalars points, and the arrows on the triangle centre are the components of the velocity vector points.}
    \label{fig:Bgrid}
\end{figure}

In addition to the B-grid discretization, FESOM's grid design plays a crucial role in computing the operators necessary for FESOM's horizontal discretization. It creates a dual cell by connecting the triangles' barycentre with its edge midpoint, creating a cell with 10 to 12 edges (Figure \ref{fig:Bgrid}).

\subsubsection{Interpolation operators}

This grid allows computing the operators by only interpolating the height field at the edges when needed to compute the gradient at the cells' barycentre. Given an edge $e$, with vertices $\widehat{K},\widehat{L} \in \partial e$, then the interpolation is defined as:

\begin{equation}
    \label{eq:BscaKtoe}
    \widetilde{h}^e = \frac{1}{2}(h_{\widehat{K}} + h_{\widehat{L}}),
\end{equation}
thus achieving a second order interpolation on the edge.

FESOM's horizontal momentum discretization is provided with three alternative computations of the momentum equations: two in its flux advective equation form, one computed at the centre of mass of the triangular cell and the other computed at the vertex, and one in a vector-invariant form, which is computed at the vertices of the grid. The two forms computed at the vertices would thus require to be interpolated at the centre of mass of the triangle with \eqref{eq:BscaKtoe}. It is also argued that the use of the flux advective form of the equation provides a small internal diffusion on the system \citep{Danilovetal2015}. However, there is a surprising lack of published work comparing these forms, indicating a need for a more in-depth research in the future. In this work, in order to ensure a fair comparison with the other schemes, we chose to compute this discretization using the vector invariant form of the equation.

\subsection{MPAS-O (C-grid)}

MPAS, an ESM from the Climate, Ocean and Sea Ice Modelling (COSIM) and National Center for Atmospheric Research (NCAR), comprises atmospheric, ocean, and ice components \citep{Ringleretal2010, Skamarocketal2012, Hoffmanetal2018, Turneretal2022}. The oceanic component has been shown capable of accurately representing geophysical flows on meshes with a large variation of resolution \citep{Ringleretal2013}.

The horizontal discretization of the dynamical core of MPAS was developed for arbitrarily sided C-grid polygons \citep{Thuburnetal2009, Ringleretal2010}. It is inspired by the Arakawa and Lamb's scheme \citep{ArakawaLamb1981}, which is capable of providing some conservative properties, such as total energy and potential vorticity, while also providing reliable simulations for these arbitrary grid structures without a breakdown of the time-integrated solutions, which has previously affected schemes using a quasi-hexagonal mesh \citep{StaniforthThuburn2012}.

Although this scheme could potentially be used for any arbitrarily sided polygonal mesh, the icosahedral based hexagonal grid was shown to provide the most accurate and well-behaved solutions \citep{Welleretal2012}. For example, analysis of this discretization has shown that the scheme can achieve at most first order accuracy for most of the operators, but a stagnation or divergent accuracy for others \citep{Peixoto2016}. Despite this, the model's noise is well controlled, while also maintaining its geostrophic modes with zero-frequency \citep{Welleretal2012, Peixoto2016}. 

\begin{figure}[h]
    \centering
    \includegraphics[width=.5\linewidth]{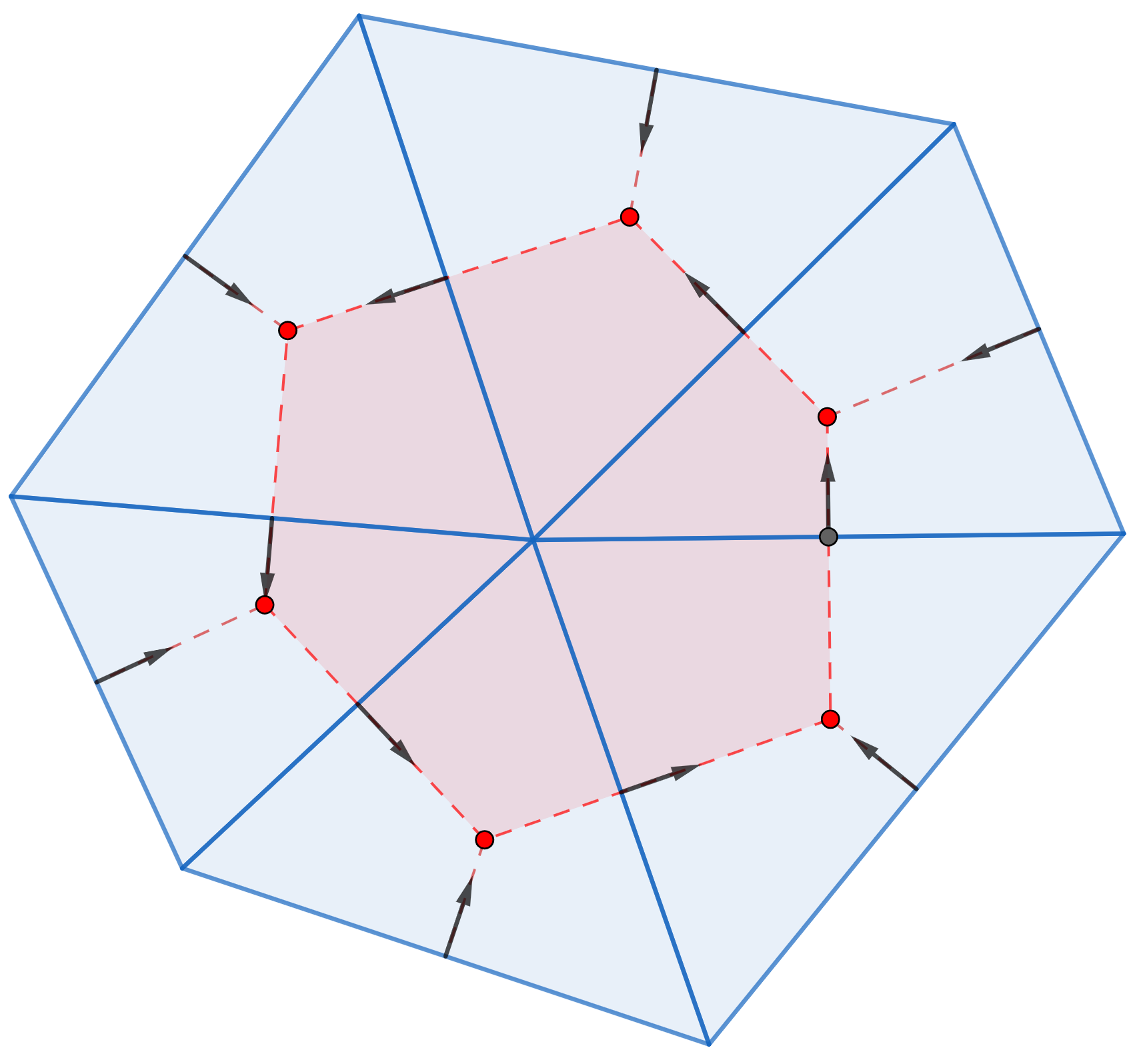}
    \caption{C-grid cell structure. Red circles on the vertices are the height scalar points, and the arrow on the edge midpoint is the decomposed velocity vector field.}
    \label{fig:Cgrid}
\end{figure}

On this C-grid discretization (Figure \ref{fig:Cgrid}), the velocity vector field is decomposed on the edges of our primal grid (triangular cells), where these velocities are normal to the dual grid (pentagonal or hexagonal cell), while the height field is collocated at the vertices of the grid. This minimizes the use of interpolating variables on this scheme. The only interpolation used is to calculate the perpendicular velocity and the kinetic energy, which will be better discussed in the following sections.

\subsection{ICON-O (C-grid)}

The ICON numerical model is a joint project between the German Weather Service and the Max Planck Institute for Meteorology and consists of atmosphere, ocean (including biogeochemistry), land, and ice components \citep{Giorgettaetal2018,Korn2017,Jungclausetal2022}. The ICON modelling team was not only able to successfully provide an accurate simulation of geophysical flow, but also provided evidence that their model is within reach to accurately simulate ocean submesoscale flow \citep{Hoheneggeretal2023}.

In the particular case of ICON's oceanic component, i.e. ICON-O, its horizontal discretization of the dynamical core is based on the mimetic methods approach, which is a practical way to discretize PDEs while taking into account fundamental properties of these equations \citep{Brezzi2014}. This philosophy, in theory, could allow for ICON depending on the truncation time to achieve the conservation of total energy, relative and potential vorticity, and potential enstrophy to some order of accuracy.

To accomplish these conservation properties under the mimetic methods, ICON-O uses the concept of admissible reconstructions $(\mathcal{P},\hat{\mathcal{P}},\hat{\mathcal{P}}^{\dagger})$ \citep{KornLinardakis2018}. These are in charge of connecting variables at different points, acting as interpolation and reduction operations. They, i.e. the admissible reconstructions, are required to have some properties, such as providing unique and first-order accurate fluxes, having its nullspace coinciding with the space of divergence noise, and conserving the aforementioned properties. However, in order to achieve these properties, it is required to compute the inverse of the resulting mass matrix on the velocity equation for each timestep. To avoid the additional computational cost, we, therefore, used the matrix lumping approach, i.e. assumed that the inverse of the mass matrix is the identity matrix. It was shown that this approach does not significantly impact the simulations of the model, nor it does significantly impact the energy conservation \citep{KornDanilov2017, KornLinardakis2018}.

\subsubsection{Interpolating operators}

Operationally, ICON-O model uses the Perot operator. This function reconstructs the velocity field components of the edge midpoint to the triangle centre ($P=\mathcal{P}$), and subsequently project these reconstructed vectors to their original position at the edge midpoint ($P^T\mathcal{P}$) \citep{Perot2000}:
\begin{align}
    Pu_K &= \frac{1}{K}\sum_{e\in\partial K} \vert e\vert u_e\mathbf{n}_e, \\
    P^Tu_e &= \frac{1}{\vert\hat{e}\vert}\sum_{K\in\partial \hat{e}} d_{e,K}u_K\cdot\mathbf{n}_e.
\end{align}
The combination of operators is denoted as $M=P^TP$ and is key to compute the operators of the shallow water equations. This mapping, $M$, was found to filter the divergence noise of triangles without losing the aforementioned physical properties \citep{KornDanilov2017, Korn2017, KornLinardakis2018}. However, the operator has the potential to smooth high wavenumber phenomena \citep{KornDanilov2017}.

Additionally, there is also a set of operators that reconstructs the vector velocity field into the vertices of the grid ($\widehat{P}=\hat{\mathcal{P}}$) and reduce it back into the edge midpoints ($\widehat{P}^{\dagger}=\hat{\mathcal{P}}^{\dagger}$). This sequence is defined as:
\begin{align}
    \widehat{P}u_{\hat{K}} &= \frac{1}{\vert\hat{K}\vert}\sum_{e\in\partial\hat{K}} \vert\hat{e}\vert u_e\mathbf{e}\times\mathbf{n}_{\hat{e}}, \\
    \widehat{P}^\dagger u_{e} &= \frac{1}{\vert\hat{e}\vert}\sum_{\hat{K}\in\partial e} d_{e,\hat{K}}u_{\hat{K}}\cdot\mathbf{n}_e.
\end{align}
Thus, the sequence $\widehat{M}=\widehat{P}^{\dagger}\widehat{P}$ allows us to compute the Coriolis term of the shallow water equations. This dual operator has shown to provide a non-zero spurious frequency geostrophic modes, which have been shown to create numerical waves in the system \citep{Peixoto2016}, and could potentially be damaging to the stability of the scheme \citep{Peixotoetal2018}. However, due to the filtering property of the operator $M$, these modes could be removed from the simulation due to their filtering property on the grid scale.

\begin{table*}[]
    \centering
    \begin{tabular}{c c c c c}
        \hline
          & Institution & Staggering & Components & Conservation\\
        \hline
         NICAM & AORI, JAMSTEC, AICS & A-grid & Atm & TE\\
         FESOM & AWI & B-grid & Oc & TE\\
         MPAS & COSIM, NCAR & C-grid & Atm/Oc/Ice & PV, TE\\
         ICON & DWD, Max-Planck & C-grid & Atm/Oc/Land/Ice & KE, TE, PV, Enst\\
        \hline
    \end{tabular}
    \caption{Summary of the main models to be compared with their respective components: Ocean (Oc), Atmosphere (Atm), Ice Dynamics (Ice) or Land; and their conservation properties: Total energy (TE), Kinetic Energy (KE), Potential vorticity (PV), and Enstrophy (Enst). }
    \label{tab:modelsum}
\end{table*}

\section{Accuracy of the Discrete Operators}
\label{sec:IdeaTestCase}

We aim to analyse the truncation errors of each operator from Nonlinear Shallow Water Equations \eqref{eq:NLSWE}. To achieve this we evaluate two different test cases: The first follows from \cite{HeikesRandall1995} and \cite{Tomitaetal2001}, henceforth Test Case 0 or TC0, where for $\alpha,\ \beta$ defined as:
\begin{align*}
    \alpha &= \sin\phi \\
    \beta &= \cos(m\phi)\cos^4(n\theta),
\end{align*}
where $\phi$ and $\theta$ are the longitude and latitude, respectively, then $\mathbf{u}$ and $h$ are defined:
\begin{align}
    \mathbf{u} &= \alpha\nabla\beta \\
    h &= \beta.
\end{align}
We consider in our analysis $m=n=1$, since it is a smooth particular smooth case with both non-zero vector components, which allows us to evaluate the accuracy of the operators and compare with the literature.

A second case is the Nonlinear Geostrophic testcase, henceforth Test case 1 or TC1, from the toolkit set of \cite{Williamson1992}. $\mathbf{u}$ and $h$ are defined as:
\begin{align}
    gh &= gb_0-h_0\sin^2 \theta \label{eq:hng}\\
    u &= u_0\cos\theta,
\end{align}
where $gb_0=2.94\times 10^4 \text{ m}^2\text{s}^{-2}$, $h_0=a\Omega u_0+u_0^2/2\text{ m}^2\text{s}^{-2}$, $u_0 = 2\pi a/(12\text{ days}) \text{ m}\text{s}^{-1}$, $g=9.81\text{ m}\text{s}^{-2}$ is the acceleration of gravity, $a=6.371\times 10^{6}\text{ m}$ is the radius, and $\Omega = 2\pi/86400\text{ s}^{-1}$ is the angular frequency of earth.

Additionally, in order to compare our results, we define the errors in our domain as $\Delta f = f_r - f_r^{\text{ref}}$, where $f_r$ and $f_r^{ref}$ is the computed and reference function, respectively, for a mesh element $r$ of the domain. Thus, the maximum and second error norm may be defined as:
\begin{align}
    L_{\infty} &= \frac{\max_r \vert \Delta f_r\vert}
    {\max_f \vert f_r^{\text{ref}}\vert} \\
    L_2 &= \sqrt{\frac{S(\Delta f^2)}{S((f^{\text{ref}})^2)}}
\end{align}
where $S(f) = \sum_{r\in\Omega}\Delta fA_r/\sum_{f\in\Omega}A_r$, and $A_r$ is the area of the element, e.g. $A_e$ for the edge, $\vert K \vert$ for triangles, or $\vert\hat{K} \vert$ for the dual cell.

\subsection{Divergence}

The divergence operator, part of the mass equation, can be defined from the Divergence Theorem. Following it, we can provide a general formula for its discretized version as:
\begin{equation}
    (\nabla\cdot\mathbf{u})_i \approx (\textbf{div }u)_i = \frac{1}{\vert F \vert}\sum_{e\in\partial F}\vert e \vert\mathbf{u}\cdot\mathbf{n}_en_{e,F},
\end{equation}
where $F$ is a cell with barycentre $i$ and edges $e\in\partial F$, $n_{e,F}=\{1,-1\}$ is a signed valued aimed to orient the normal velocity $\mathbf{u}\cdot\mathbf{n}_e$ away from the element $F$.

In order to compute the divergence field, we note that both the A-grid and B-grid schemes compute divergence field at the dual cells (vertices). For the former scheme, we require an interpolation of both the scalar height, \eqref{eq:AScavtoK} and \eqref{eq:AScaKtoe}, and vector velocity fields at the dual edge midpoint, in order to compute the divergence at the dual cell, i.e. quasi-hexagonal cell. In the case of the latter scheme, we only require the interpolation of the scalar height field at the primal edge midpoint \eqref{eq:BscaKtoe}, in order to compute the same divergence field at the primal cell.

In the case of the C-grid, there is a substantial difference between the computation of both schemes. MPAS interpolates the scalar height field at the primal edges, similar to B-grid, while ICON uses admissible reconstruction operators of the form $P^ThPu$ to compute the operator.

These differences on the schemes are reflected in our results (Figure \ref{fig:operatorserror}.div). The A-grid for the TC0 testcase displayed an error convergence with an initial rate of second order up to the sixth refinement (g$_6$). On finer grids, for the $L_\infty$, this scheme has slowed down to first order, while on second order, the scheme remained converging up to second order rate. On the TC1, a more consistent convergence rate was observed, on the L$_\infty$ and L$_2$, the scheme has displayed a first and second order convergence rate. On other grids, in particular the standard and Spring Dynamics, the A-grid has shown to achieve at least a first order convergence rate \citep{Tomitaetal2001}. Although a direct comparison cannot be provided, since our testcases differ, the scheme on an SCVT has apparently shown to provide a comparable convergence rate to the intended optimized grid on either the L$_\infty$ or the L$_2$ norm.

Regarding both C-grid schemes, we observe a similar behaviour in the computed operator. In particular, neither scheme displays an increase in accuracy of the divergence field on the $L_\infty$. For the case of ICON, this result has been previously observed in a similar work by \cite{KornLinardakis2018}. It was also shown that the \textit{naive} approach to calculate the divergence field still retained a first order increase in accuracy, implying that the main culprit of this inability to increase the accuracy likely lies on Perot's operator itself (Table 4 of \cite{KornLinardakis2018}). The authors have not provided a geometrical analysis of their non-uniform grid, but we note that the SCVT grid share some similarities with the standard grid, such as the non convergence of the distance between the primal and dual edge midpoints, which likely has a deleterious effect on the accuracy of the operator. However, on the L$_2$, the scheme was able to reach at least a first order convergence rate on both testcases.

On the case of MPAS, the inability to provide a decrease in error with grid has been discussed in \cite{Peixoto2016}. It is reasoned that since the computation of the divergence is not based on velocities from the Voronoi edge midpoints, the discretization is inconsistent, and a first order convergence is not guaranteed. In contrast, on the $L_2$, MPAS was able to reach a second order rate up to g$_4$, but the speed of convergence slows down to first order on TC0, while on TC1 the second order rate is maintained throughout grid refinements.

Finally, B-grid has provided consistent accuracy throughout each testcase. We observed a first and second convergence rate for $L_\infty$ and $L_2$, respectively, for both testcases. A decrease is observed on TC0, however, this decrease is likely associated with the error approaching the machine truncation error.

When comparing the errors of the schemes, we note that both A- and B-grid schemes display a decrease in speed of accuracy convergence as the grid is refined, with the latter scheme displaying the smallest errors on most of the tested cases and error norms. Additionally, despite ICON providing convergence on some tests, the scheme displays the largest errors of all tested schemes. It is likely that the smaller stencil used in ICON's divergence computation play a role in these larger errors. Another contribution is potentially related to Perot's operator, whose interpolation could act as smoothing the velocity field.

Overall, we note that the structure of the mesh, regarding cell geometry (primal or dual cell) and distortion, plays a contributing factor on approximating the divergence field on all schemes. Both C-grid schemes, in particular, seemed to be the most vulnerable to the grid. In contrast, B-grid's consistency in its accuracy apparently seems to be the least vulnerable to the increase in the distortion of the grid.

\subsection{Gradient}

The gradient operator, from the momentum equation, is a vector field, whose vector points itself to the steepest regions of the original field. The schemes provide different discretizations for this operator:
\begin{equation}
    \nabla h \approx \textbf{grad }h =
    \begin{cases}        
        \sum_{e\in\partial F} h\vert e\vert \mathbf{n}_e & \text{A- and B-grid},\\
        \frac{1}{\vert e\vert}\sum_{i\in\partial e} h n_{e} & \text{C-grid}.
        \end{cases}
\end{equation}
A- and B-grid's schemes provide a complete vector field on our domain by computing the average gradient within the centre of the respective cell $F$. The C-grid, on the other hand, computes the gradient with respect to the normal vector $n_e$ by computing the difference between the values of the cell neighbouring the edge $e$. In that regard, the C-grid computation can be perceived as a gradient in the direction of $\mathbf{n}_e$.

In relation to the mesh, the A-grid scheme is computed at the vertices of the mesh, while the B-grid is computed at the barycentre of the triangular cells. On the other hand, both C-grid schemes are computed on the primal edge midpoint of our mesh. However, the MPAS scheme considers the neighbouring vertices to compute the gradient, while ICON considers the neighbouring triangles.

As in the divergence approximation, these differences in computation are as well reflected in our results (Figure \ref{fig:operatorserror}.grad). The A-grid displays for coarser grids a fast convergence rate (second order rate), up to $g_5$, for both testcases. For finer grids, the $L_\infty$ the decrease in error slows down to a first order convergence, but with the $L_2$ the convergence rate remains consistent. The analysis made by \cite{Tomitaetal2001} have showed that their grid is capable of displaying a second order error convergence. We again note that although we cannot directly compare our results, due to the differences in testcases used, our results show a comparable error convergence with the authors with the SCVT optimized grid.

Similarly, the B-grid scheme shows a consistent decrease in error on all norms and testcases, similar to the divergence operator results. However, it displays only a first order convergence rate, in contrast to the second order on the divergence operator. The computation of the gradient on the B-grid is analogue to the divergence computation in ICON, therefore a similar argument follows, explaining that the expected convergence rate of such a scheme being a first order.

Comparably, MPAS also displays a consistent convergence rate, but in this case this scheme achieves a second order rate on all norms and testcases. Since the edge midpoint is situated, by definition, at the midpoint between the neighbouring vertices, the discretization is analogue to a centred difference scheme used in traditional quadrilateral grids. Therefore, we can properly achieve a second order convergence rate. The same argument is provided in \cite{Peixoto2016}, however the author also argues that when we consider the computation of the gradient of the kinetic energy we do not only reach a convergence rate, but our error diverges with grid refinement. The author reasons that the error of kinetic energy is of zeroth order (to be discussed further), and, thus, its gradient diverges.

On the other hand, the ICON's scheme gradient error displays a near second order convergence rate for coarser grids on the $L_\infty$ norm of the TC0, but this error slows down for further refinements. On the TC1 testcase, the rate of convergence on $L_\infty$ is consistent in first order. However, at the $L_2$ norm, the scheme has an accuracy of near second order with magnitude similar to that of MPAS.

Finally, we can then draw a comparison from all schemes. The B-grid has displayed the largest errors in magnitude and was the only scheme to achieve a low first order convergence on the $L_2$. The A-grid $L_\infty$ displays a similar error magnitude and  behaviour in convergence with ICON. MPAS has shown the lowest errors among all schemes, and, in the $L_2$, displayed a comparable magnitude and convergence behaviour with ICON.

Overall, we again observe an impact of the grid structure on our schemes, however, this impact is not as damaging as found in the divergence computation. The directional derivative of MPAS makes it easier to achieve a consistent increase in accuracy, and the mismatch between the edge midpoints, has thwarted ICON's convergence rate. Despite this, the scheme still retained a first order convergence rate.

\subsection{Curl}
The curl operator, part of the vector invariant form of the shallow water velocity equation, is connected to the Coriolis Term. This term requires a careful discretization to allow for Coriolis energy conservation. This operator, in its continuous form,  is defined from Stokes Theorem. Its Finite Volume discretization follows from this theorem and a general formulation for all our schemes can be defined as:
\begin{equation}
    \nabla\times\mathbf{u}_i \approx \vert F\vert\text{vort }u_i= \sum_{i\in\partial F} \vert e'\vert \mathbf{u}_i\cdot\mathbf{t}_{e'}t_{e,F},
\end{equation}
for any $F$ cell with edges $e'$, tangent vector $\mathbf{t}_{e'}$, and $t_{e,F}=\{1,-1\}$ is a signed value guaranteeing that the unit tangent vector is counterclockwise on the cell.

For each scheme, the both A-grid, and B-grid computes the vorticity field on the vertices of the mesh. Since, for the B-grid, the shallow water velocity equation requires the points at the barycentre of the triangle cell, we interpolate the vorticity from the vertices to the barycentre. For the both C-grid schemes, MPAS computes this operator at the circumcentre of the cell, while ICON computes at the vertices, in duality with the divergence operator.

In this context, similarities are observed  with the divergence operator. For example, the A-grid convergence rate for both norms and testcases, reach the same order as the divergence operator. On the TC0 testcase, however, throughout all grid refinements the error retain a first order, unlike the divergence operator, which begins with a second order and slows down to a first order. Additionally, on the TC1 testcase, we observe that the vorticity error displays a second order convergence up to g$_4$ and slows down to first order, unlike the divergence operator (Figure \ref{fig:operatorserror}.Vort).

Similarly, the B-grid scheme displays the same behaviour as in the divergence operator. It displays a first order convergence rate on $L_{\infty}$ and a rate of second order for $L_2$ on both testcases.

In contrast, both C-grid schemes display a different behaviour from the divergence operator. MPAS shows a consistent first order convergence rate for both norms on both testcases. Given that this computation is computed on the dual cell centre (red polygon in Figure \ref{fig:Cgrid}), i.e. pentagon or hexagon, we can then achieve a higher accuracy rate of around second order.

ICON, on the other hand, displays a zeroth order convergence on $L_\infty$ for the TC0 testcase. This is likely due to the mismatch of edge midpoints, similar to MPAS's divergence operator. However, on this norm for TC1, the error converges on a first order rate. This difference implies that different testcases will potentially impact the error. On this particular case, we note that the meridional velocity is not present on TC1, which may facilitate the computation of the vorticity. This result is also seen on $L_2$, while for TC0, the norm converge in first order, for TC1, it converges in second order.

In comparison, we observe that ICON is the only scheme that has trouble in increasing its accuracy when approximating the vorticity operator. In addition, both A- and B-grid schemes were the only to display a second order error rate on the L$_2$ for both schemes. Although MPAS also has shown an overall convergence, in contrast to ICON, it still has shown a larger error for TC0's $L_2$ norm and both norms of TC1.

Overall, there are similarities on the error behaviour between both vorticity and divergence scheme due to its similar concepts underlying the discretization. In that regard, we also observe an impact of the grid structure and the testcase used on the accuracy of the vorticity approximation.

\subsection{Kinetic Energy}

Similar to the vorticity operator, the kinetic energy is part of the vector invariant form of the velocity equation of the shallow water, whose gradient will then be computed. The kinetic energy is defined as:
\begin{align*}
    E_k = \frac{1}{2}\vert\mathbf{u}\vert^2.
\end{align*}
The computation of this operator on both A- and B-grid schemes is straightforward, since the vector velocity field is complete on each vertex and barycentre, respectively, of the mesh. However, for the C-grid schemes the vector field is decomposed on the edges of the mesh, therefore require a reconstruction in order to approximate the value of the kinetic energy field. In the particular case of MPAS and ICON, it is difficult to provide a general formula, therefore we individually define:
\begin{align}
    E_k^{(\text{MPAS})} &= \frac{1}{2\vert \hat{K}\vert}\sum_{e\in\partial\hat{K}}
        \frac{\vert e\vert \vert\hat{e}\vert}{2} u_e^2, \\
    E_k^{(\text{ICON})} &= \frac{\vert Pu \vert^2}{2}.
\end{align}
Both schemes provide some form of interpolation of the velocity on the cell centre, dual for MPAS, primal for ICON. It is observed on this computation that MPAS's and ICON's weights are shown to be: $\vert e \vert\vert\hat{e}\vert/2$, and $\vert e \vert d_{e,K}$, where again $d_{e,K}$ is the distance between the edge midpoint $e$ and circumcentre $K$. We note that for equilateral triangles $d_{e,K}=\vert\hat{e}\vert/2$. Another note is that MPAS computes the square of the component of the velocity and then interpolates the resultant on the cell centre, while ICON interpolates the complete vector velocity field on the cell centre, and then computes the kinetic energy.

These difference in computation are reflected on the error of the field (Figure \ref{fig:operatorserror}.Ek). On MPAS scheme, we see that for both testcases it does not converge on $L_\infty$. This result was discussed by \cite{Peixoto2016}, as being an inconsistent formulation of the kinetic energy on the SCVT. Part of this inconsistency could partly be due to the computation of the kinetic energy on a single velocity component, as previously mentioned. Despite this, on $L_2$, MPAS display a second order convergence on TC0, on coarser grids, but it slows down to first order on finer grids. Similarly, on TC1, MPAS displays a first order rate, but throughout all grids.

ICON, in contrast, show a consistent convergence rate on both norms of first order on TC0 and second order on TC1. It can also be observed that, except for TC0's $L_2$, ICON's error is substantially lower than MPAS. ICON's Perot operator interpolation allows for a higher convergence, in comparison with MPAS, in part due to the vector velocity field interpolated on the cell circumcentre prior to the computation of the kinetic energy.

Overall, both C-grid computations display very distinct error behaviour. On this grid, although on both schemes the kinetic energy formulation allows for energy conservation, MPAS is unable to provide a consistent formulation of the operator. In contrast, ICON is provided with its consistent through the use of its Perot operator.

\subsection{Perpendicular Velocity}

The perpendicular velocity is an important part of the Coriolis Term, which is a forcing that takes into account the non-inertial reference frame of the shallow water equations. In that case, it is important that the Coriolis term of our schemes does not input energy into the system. Similar to the kinetic energy, both the A- and B-grid schemes have their vector velocity defined on the same points, providing an exact value for the perpendicular velocity. However, since C-grid schemes do have their vector velocity decomposed on the edges of the grid, an interpolation is necessary.

This interpolation should be carefully chosen in order to retain the conservation of energy of the system.  Following the argument of \cite{Peixoto2016}, a reconstruction can be thought as a weighted composition of the neighbouring edges of the cell:
\begin{align}
    u^{\perp}_e = \sum_{e'} w_{e,e'}u_{e'}.
    \label{eq:uperpek}
\end{align}
These weights should be chosen such that this reconstruction is unique and does not provide energy to the system.

Choosing the edges $e'$ from cells that share the same edge $e$ we can define the perpendicular velocity as:
\begin{align}
    u^{\perp}_e = a_{e,F_1}u^{\perp}_{e,F_1} + a_{e,F_2}u^{\perp}_{e,F_2},
    \label{eq:uperpe}
\end{align}
where $a_{e,F_n}$ are the weights with respect to the cell $F_n$. This formulation is capable of achieving a unique solution on the edge.

In the case of MPAS's vector interpolation, we define the weights $w_{e,e'}$ as:
\begin{align*}
    w_{e,e'}=c_{e,K}\frac{\vert e'\vert}{\vert\hat{e}\vert}\left(\frac{1}{2}
        -\sum_{K\in\cup\partial e}\frac{A_{\widehat{K},K}}{\vert\widehat{K}\vert}\right)n_{e',\hat{K}},
\end{align*}
where $c_{e,\hat{K}}$ and $n_{e',K}$ are sign corrections that guarantee the vector tangent vector is anticlockwise on the for the cell $\hat{K}$ and that the norm vector $n_{e'}$ point outwards of the cell $\hat{K}$; and $A_{\hat{K},K}$ is the sectional area of the triangle cell $K$ formed by the vertex $\hat{K}$ and the neighbouring edges of the circumcentre $K$ in respect to the vertex. Using these weights on \eqref{eq:uperpek}, we can compute $u^\perp_{e,K}$. In order to provide a unique reconstruction on edge $e$ we let $a_{e,K} = a_{e,L} = 1$ on \eqref{eq:uperpe}.

\begin{figure*}[ht]
    \centering
    \includegraphics[width=\linewidth]{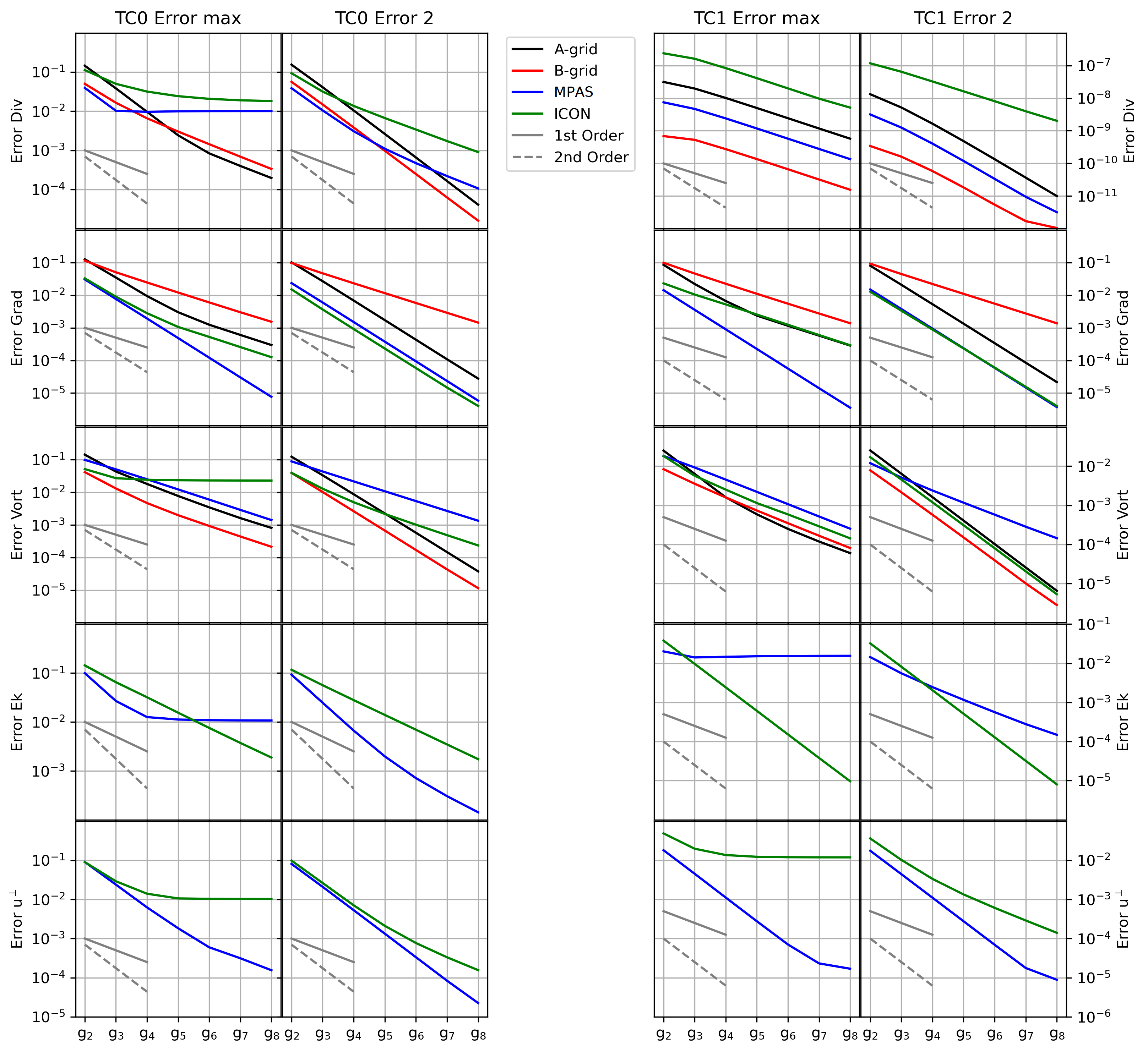}
    \caption{TC0 (first and second row panels) and TC1 (third and fourth row panels) operators $L_\infty$ (first and third panels) and $L_2$ (second and fourth panels) error norms for the A-grid (black lines), B-grid (red lines), MPAS (blue lines), and ICON (green lines).}
    \label{fig:operatorserror}
\end{figure*}

In the case of ICON's scheme, we use the interpolation $\hat{P}^T\omega\hat{P}u$. In this case $\hat{P}u_{\hat{K}} = u^{\perp}_{\hat{K}}$, so the weights are defined as:
\begin{align*}
    w_{e,e'} = w_{\hat{e},\hat{K}} = \frac{\vert\hat{e}\vert d_{\hat{e},\hat{K}}}{\vert\hat{K}\vert},
\end{align*}
giving a unique reconstruction on the centre of the dual cell $\hat{K}$. In order to reduce it back to the edge, we do $a_{e,\hat{K}} = d_{e,\hat{K}}/\vert e \vert$. We note that this set of operators allows not only the energy conservation, but also potential enstrophy \citep{KornDanilov2017,KornLinardakis2018}. We recall, however, that this operator has the potential of producing non-zero frequency geostrophic modes \citep{Peixoto2016}.

Our results show that MPAS displays a second order convergence rate on $L_\infty$ up to g$_6$ on TC0, but decrease to a first order for finer grids (Figure \ref{fig:operatorserror}.u$\perp$). On $L_2$, it shows a second order throughout all refinement. Similarly, on TC1, it also shows a second order rate up to g$_7$, but decrease near first order to g$_8$. A similar result is obtained for $L_2$. This result is similar to \cite{Peixoto2016} showing that MPAS achieves at most a first order convergence rate on the $L_\infty$.

\FloatBarrier
\section{Shallow Water Time Integration}
\label{subsec:SWTI}

The time integration of the shallow water equations provides us knowledge about the behaviour and limitations of the model throughout time. In order to gather this understanding, in this section we will put the schemes under a battery of tests. For the purpose of these tests, we chose to use a simple Runge-Kutta (RK44) operator, with 50 seconds timestep for all schemes and grids. Such choices are enough to ensure that the temporal errors are minimal and that the dominating error comes from the spatial discretization. We note that although both C-grid schemes may not require a stabilization term, since their error are expected to be well controlled, both A- and B-grid schemes could excite errors that would potentially destabilize the model. It is possible to use a harmonic ($\nabla^2\mathbf{u}$) or biharmonic ($\nabla^4\mathbf{u}$) term to provide stability of the scheme. In order to be more scale selective and avoid damping physical waves of our simulations we chose to use only the biharmonic, and as it was shown by the original authors of A- and B-grid schemes \citep{Tomitaetal2001,Danilovetal2017} the biharmonic term is enough to provide the necessary stability.

Therefore, the stabilizing operator can be regarded as a composition of Laplace diffusion operators, i.e. $\nabla^4\mathbf{u} = \Delta\Delta\mathbf{u}$. To compute the Laplace diffusion operator, both A- and B-grid schemes are equipped with different approaches in its computation. For the former scheme, the Laplace operator is defined as:
\begin{equation}
    \label{eq:diffA}
    \Delta\mathbf{u} = \nabla\cdot\nabla\mathbf{u}.
\end{equation}
Thus, we can approximate the Laplacian operator by $\Delta\mathbf{u}\approx\textbf{div }\textbf{grad }\mathbf{u}$, using the operators defined in the previous section.

On the other hand, the B-grid scheme, computes the harmonic diffusion for a cell $K$ as:
\begin{equation}
    \label{eq:diffB}
    \Delta\mathbf{u} \approx \frac{1}{\vert K\vert}\sum_L \frac{\vert e\vert}{\vert\hat{e}\vert}(\mathbf{u}_L-\mathbf{u}_K),
\end{equation}
where $L$ are all the triangles neighbouring the cell $K$. For the tested schemes, we used the biharmonic coefficient defined in Table \ref{tab:biharcoef}. Our coefficients are much higher than found in literature \citep{Tomitaetal2001, Danilovetal2017, Majewskietal2002, JablonowskiWilliamson2011}, however both A- and B-grid schemes differ in their discretization and the A-grid scheme is found susceptible to numerical oscillations depending on the choice of grid \citep{Tomitaetal2001}. Therefore, by choosing an intense coefficient, we guarantee that numerical waves will not participate in the comparison of our results.

\begin{table}[]
    \centering
    \begin{tabular}{c c}
        \hline
         & A-grid/B-grid (m$^2$s$^{-1}$)\\
        \hline
         g$_2$ & 10$^{22}$\\
         g$_3$ & 10$^{20}$\\
         g$_4$ & 10$^{19}$\\
         g$_5$ & 10$^{18}$\\
         g$_6$ & 10$^{17}$\\
         g$_7$ & 10$^{16}$\\
         g$_8$ & 10$^{15}$\\
        \hline
    \end{tabular}
    \caption{Biharmonic coefficient used for stabilizing the shallow water schemes.}
    \label{tab:biharcoef}
\end{table}

All schemes will then be evaluated. Firstly, we provide an accuracy analysis of the integrated height and vector velocity fields (Section \ref{subsec:tiav}). Then, we evaluate the linear mode analysis of our schemes (Section \ref{subsec:LNM}). Thirdly, we evaluate the scheme's capacity in maintaining its geostrophic balance (Section \ref{subsec:LBF}). Finally, we evaluate the behaviour of each scheme under a barotropic instability, which is an initial condition that accentuate the nonlinear terms of our schemes (Section \ref{subsec:BI}).

\subsection{Time integrated accuracy of variables}
\label{subsec:tiav}

\begin{figure*}
    \centering
    \includegraphics[width=\linewidth]{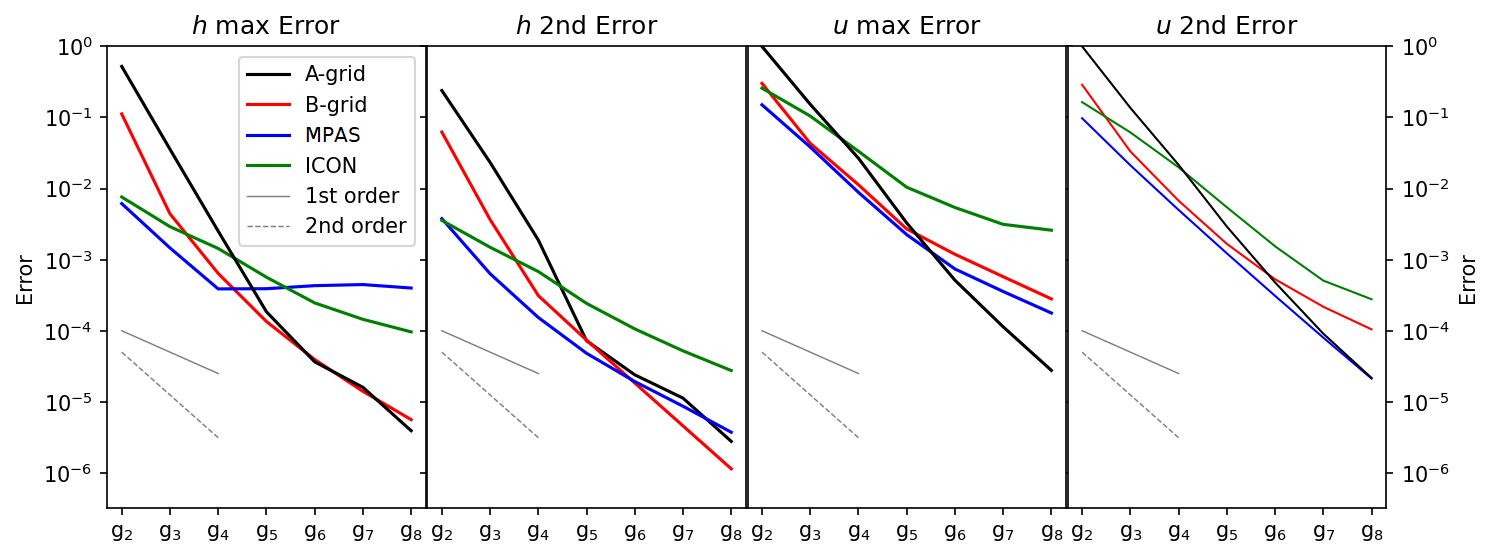}
    \caption{$h$ and $u$ error after 15 days.}
    \label{fig:varTC0}
\end{figure*}

Our results demonstrate that both A- and B-grid schemes exhibit improvements in accuracy close to second order for both norms of the height field variable (Figure \ref{fig:varTC0}). However, for the vector velocity field, the values differ. For $L_\infty$, A-grid is shown to converge near second order, while B-grid, which displays a near second order convergence for coarser grids (up until g$_5$), only shows a first order for the finer grids. Nevertheless, on $L_2$, both schemes are shown to display an accuracy increase near second order.

Regarding both C-grid schemes, both of them face problems on increasing their accuracy on $L_\infty$. MPAS does not converge on the height scalar field, but does display a first order convergence rate on $L_2$. Concerning the vector velocity field on $L_\infty$, MPAS shows a seconder order rate for coarser grids (up until g$_6$), but decrease to first order in finer grids. However, on $L_2$, MPAS displays a second order rate consistently for all refinements. This result was also observed in \cite{Peixoto2016}, and it is suggested that either the kinetic energy approximation or the divergence, might be responsible for reducing the solution's accuracy.

In contrast, ICON displays a first order convergence rate on both norms for the height scalar field. Nevertheless, the scheme does not seem to convergence on the vector velocity field for the $L_\infty$ norm. In the case of $L_2$, it displays, for coarser grids, a second order accuracy rate, but from g$_7$ to g$_8$ it slows down to a first order rate. Similar to MPAS, some operators, face challenges in converging the solution. In this scheme, the divergence, vorticity, and the perpendicular velocity do not display a convergence of the solution. It is noted that both vorticity and perpendicular velocity are critical components of the Coriolis Term of \eqref{subeq:SWvel}, potentially impacting the convergence of the vector velocity field. \cite{KornLinardakis2018} did not observe the same results. Therefore, it is likely that the grid choice is crucial for obtaining convergence on the fields.

Overall, A- and B-grid display similar errors, specially, in the height field. ICON's scheme have showed the largest errors of the tested schemes, except in the height field $L_\infty$, where MPAS did not converge. B-grid show the second-largest magnitude error, only on the vector velocity field. This is likely due to the use of the biharmonic and the notably due to the gradient operator that is defined on triangles, unlike both A-grid and MPAS, which shows similar magnitudes on $L_2$. On $L_\infty$, however, MPAS shows a larger error and lower convergence rate, in comparison to the A-grid, likely due to the aforementioned challenges.

\subsection{Linear Normal Modes}
\label{subsec:LNM}

The earth's ocean behaviour is modulated by oscillations that are mostly affected by the earth's rotation. The complete nonlinear equations are difficult to analyse to the high degree of interactions between these oscillations. However, linear analysis can be done by considering  \eqref{eq:NLSWE} the following approximations:
\begin{equation}
    \label{eq:Lswe}
    \begin{split}        
        h &= H\nabla\cdot\mathbf{u} \\
        \mathbf{u} &= -\nabla h - f\mathbf{u}^\perp,\\
    \end{split}
\end{equation}
where $H$ is a fixed constant. This system still provides a large set of inertia-gravity waves present in either the ocean or atmosphere. In order to calculate the normal modes, we follow the methodology of \cite{Welleretal2012} by considering a vector $(\mathbf{h},\mathbf{u}')^T$, where both elements, i.e. $\mathbf{h}$ and $\mathbf{u}$, are scalars, so that we have $(\mathbf{h},\mathbf{u}')^T =[h_1,h_2,\cdots,h_M,u_1,u_2,\cdots,u_N]$ for $M$ and $N$ elements of height and velocity fields, respectively. In the case of A- and B-grid, the scalar velocity is obtained by decomposing them into zonal and meridional velocity scalars, whereas for both C-grid schemes these scalar fields are obtained directly from the velocity on the edges of the grid.

We run \eqref{eq:Lswe} $M+N$ times for one timestep of $\Delta t = 10$ seconds on a g$_2$ grid, with the RK4. The initial conditions used are defined by a unit value on the j-th position of $(\mathbf{h},\mathbf{u}')^T$, i.e. for the k-th run the initial condition is defined as $(\mathbf{h}_0,\mathbf{u}_0')^T_k=[\delta^k_j]$, where $\delta^k_j$ is the Kronecker delta. We use as parameters: $gH=10^5$ m$^2$s$^{-2}$, $f=1.4584\times 10^{-4}$ s$^{-1}$ and the radius of the earth $a = 6.371\times 10^6$.

From these runs, we create a matrix $A$, where each column is the approximated solution of the initial condition provided. We, then, can calculate the eigenvalues $\lambda$ of the matrix and, consequently, obtain the frequency of the modes from $\lambda=\alpha e^{i\omega\Delta t}$, where $\omega$ is the frequency of the normal modes. We, then, order our results from lowest to maximum frequency. We will have 486 eigenvalues for the A-grid, 642 for both B-grid and MPAS, and 800 for ICON. These values correspond to the total degrees of freedom of our system. There are, in the g$_2$ grid, 162 vertices, 480 edges, and 320 triangles. For the A-grid, since both mass and vector fields are defined at the vertices, the total DOFs are three times the vertices. In the case of the B-grid, the vector field is defined at the triangles, therefore the total DOFs are the vertices plus twice the triangles. For both C-grid schemes, the vector velocity field is defined at the edges, however MPAS has the mass at the vertices, while ICON has the mass defined at the triangles. In that case, MPAS DOFs are the vertex plus edge points and ICON is the triangle points plus edge points.

\begin{figure}
    \centering
    \includegraphics[width=.7\linewidth]{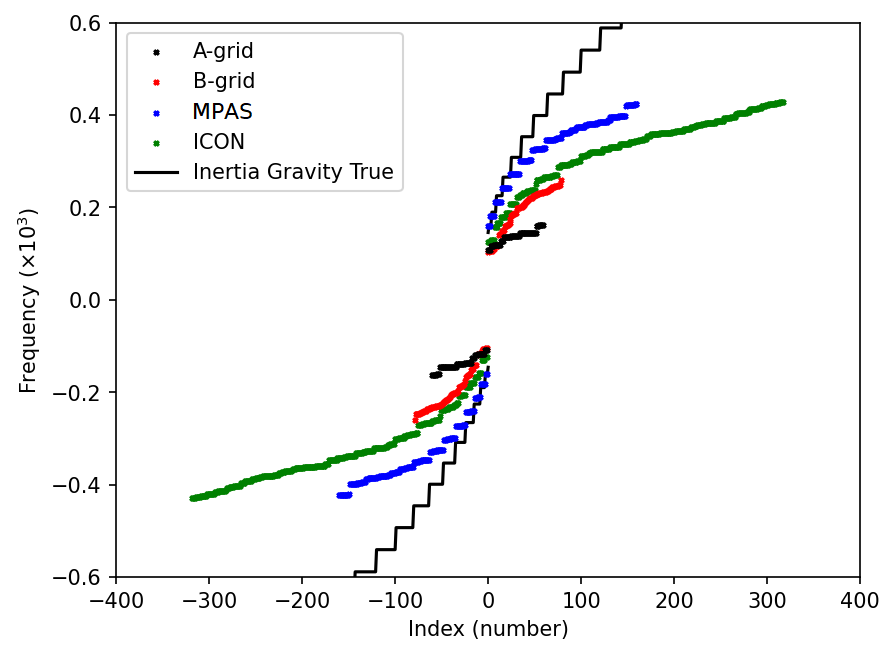}
    \caption{Linear normal modes of the considering the linear shallow water equations \eqref{eq:Lswe} on the $f$-sphere.}
    \label{fig:normmodes}
\end{figure}

The normal modes can be seen in Figure \ref{fig:normmodes}. A clear difference is observed between frequency representation on all grids. The A-grid shows the slowest representation of inertia-gravity waves, with the maximum frequency of 1.6$\times$10$^{-3}$ s$^{-1}$ s$^{-1}$ on the 119 index. On the other hand, the B-grid scheme shows higher frequencies, with a maximum on the 167 index of around 2.6$\times$10$^{-3}$ s$^{-1}$.

In contrast, a more accurate representation is obtained by both C-grid schemes. ICON shows a similar, but slightly higher frequencies, compared to the B-grid scheme. However, the highest frequency is obtained on its tail on the 635 index of around 4.2$\times$10$^{-3}$ s$^{-1}$. Conversely, MPAS displays a more accurate representation of the modal frequency with a maximum on index 320 of around 4.2 $\times$10$^{-3}$ s$^{-1}$.

Overall, our results show similar results with the traditional quadrilateral grids \citep{ArakawaLamb1977, Randall1994}. It is known that on these grids, the C-grid schemes represent modes more accurately than the either A- or B-grid schemes, but also B-grid display a higher frequency, and a more accurate representation of inertia-gravity waves, than the A-grid schemes. We highlight that the expected decrease in inertia-gravity representation from the traditional grids is not observed in our results, since we reordered our modes from least to highest frequency. Consequently, higher modes (higher wavenumbers) of both A- and B-grid schemes are not accurately displayed in our results. Despite this, our results demonstrate that the maximum represented frequency of both schemes are indeed lower than that of the C-grid schemes, following the theory.

Regarding both C-grid schemes, our results for MPAS agree with the other authors \citep{Welleretal2012, Thuburnetal2009, Peixoto2016}. In addition, we note that ICON's has a less accurate representation of the normal modes in comparison with on MPAS either on the quasi-hexagonal grid or its implementation on triangles \citep{Thuburnetal2009}. This result in ICON has already been observed \citep{KornDanilov2017}, and it is argued that the filtering property of the divergence on the mass equation might not only remove the intended noise of the triangular mesh, but also some of the higher frequency physical oscillations.

\subsection{Localized Balanced Flow}
\label{subsec:LBF}

An important testcase is to evaluate the model's capability of maintaining its geostrophically balanced state. Our TC1 testcase (Section \ref{subsec:tiav}), allowed us to test whether the models are capable of maintaining their state under small wavenumbers. However, a harder evaluation is to test whether the model have the ability to maintain its state under high wavenumber oscillations. For this reason, we used the testcase developed in \cite{Peixoto2016}. This test is particularly important for two main reasons: one of them is that the Perot's operator might not have steady geostrophic modes which may have consequences for the ICON model, the second reason is that both A- and B-grid are unable to maintain their geostrophic balanced state. We evaluate, without the stabilizing term, how all models behave under this testcase.

On that account, we define the testcase as follows:
\begin{equation}
    \begin{split}
        h&=h_0(2-\sin^n\theta) \\
        u_\phi &= \frac{-F + \sqrt{F^2+4C}}{2},
    \end{split}
\end{equation}
where $h_0$ is a constant, such that $gh_0=10^5$m$^2$s$^{-2}$, and $n=2k+2$ for any positive $k$. In our particular case, $k=160$. We also define $F$ and $C$ as:
\begin{align*}
    F&=af_0\frac{\cos\theta}{\sin\theta} \\
    C&=g_0n\sin^{n-2}(\theta)\cos^2(\theta).
\end{align*}
We will also consider the f-sphere with $f_0=1.4584\times10^{-4}$ s$^{-1}$. Finally, the grid is rotated so that the nucleus of the depression is centred at 1$^\circ$E, 3$^\circ$N.

The parameters used in this testcase will have a timestepping scheme and timestepping value as defined in section \ref{subsec:SWTI}. We will also use a g$_6$ refinement, where there are abrupt changes on the height field in a very restrict number of cells.

\begin{figure}[h]
    \centering
    \includegraphics[width=.7\linewidth]{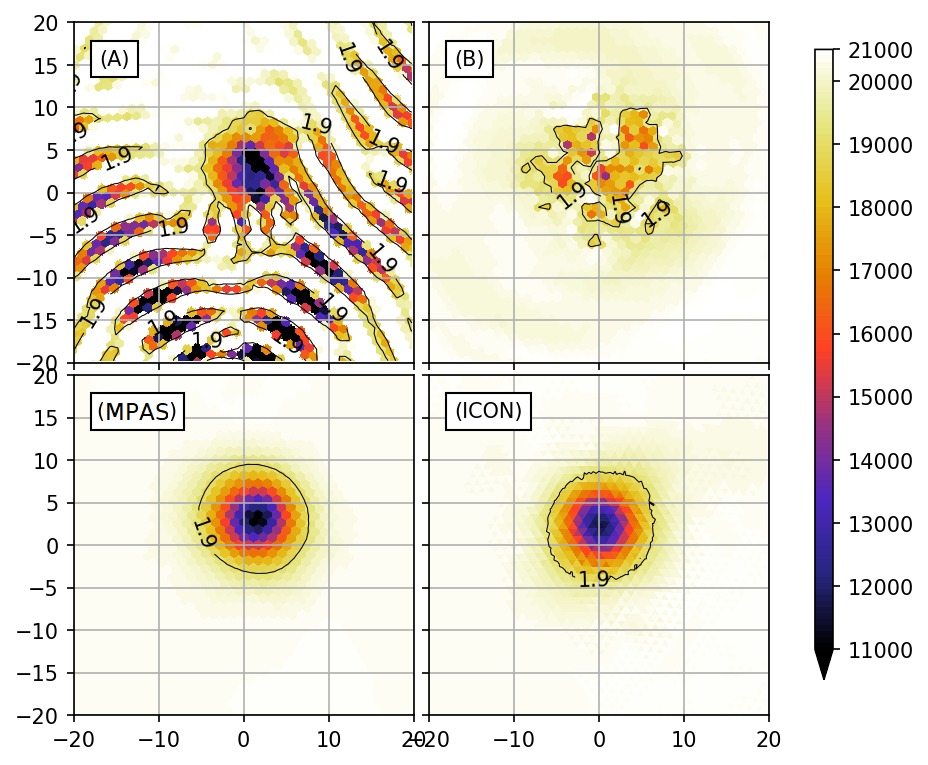}
    \caption{Height field of the different schemes for the localized balanced flow testcase without using biharmonic for both A- and B-grid schemes. Using a grid refinement g$_6$ and a timestep of 50s.}
    \label{fig:heightBalanced5Day}
\end{figure}

Our results displayed in Figure \ref{fig:heightBalanced5Day} show that both A- and B-grid, without the stabilizing term, are not capable of maintaining the geostrophic balance. For the A-grid, the numerical artefacts, emanated primarily from the pentagons of the grid, destabilize the scheme leading to an exponential growth blowing up the model around the 40 hours integration. In contrast, in the case of the B-grid scheme, there was not detected the presence of fast spurious numerical oscillations. However, the detected numerical dispersion waves were capable of breaking the down the depression up until the 24 hours after the start of the simulation.

Conversely, both C-grid schemes maintain the depression throughout the 5-day period of integration. However, in ICON's case there is a small presence of a noise on the system, but it does not seem to be enough to impact the overall solution.

Overall, the solution of A- and B-grid are impacted from their numerical oscillations. Although in the work of \cite{Yuetal2020} the A-grid is capable of integrating for a long time, the small wavelength oscillations in this testcase, generated mostly on the pentagons of the mesh, destabilize the integration, blowing up the solution. In contrast, both C-grid schemes solutions do not display damaging oscillations on the solution. MPAS's scheme and Perot's operator on the dual grid for this testcase has been observed by \cite{Peixoto2016} and observed the scheme accurately maintain their geostrophic state. We show are able to show that on the primal grid, ICON, with the use of Perot's formulation, is also able to represent the geostrophic balance state on small scale flows, despite the issues on accuracy of its operators on the SCVT (Section \ref{sec:IdeaTestCase} and \ref{subsec:tiav}).

\subsection{Barotropic Instability}
\label{subsec:BI}

Previous testcases aimed in studying the fluid flow under highly controlled experiments, in order to evaluate their accuracies, linear normal modes, and balanced state flow. However, the highly energetic and chaotic nature of the ocean require a more realistic testcase, such a fluid flow instability.
\begin{equation}
    \label{eq:Ggeos}
    \begin{split}
        u &=
        \begin{cases}
            \frac{u_{\text{max}}}{e_n}\exp\left[\frac{1}{(\phi-\phi_0)(\phi-\phi_1)}\right] & \phi_0<\phi<\phi_1 \\
            0 & (\phi-\phi_0)(\phi-\phi_1) >0
        \end{cases} \\
        gh(\phi) &= gh_0-\int^{\phi}_{-\pi/2}au(\phi')\left[f+\frac{\tan(\phi')}{a}u(\phi')\right]d\phi'.
    \end{split}
\end{equation}
where $u_{\text{max}}= 80 \text{ms}^{-1}$, $\phi_0 = \pi/7$, $\phi_0 = \pi/2-\phi_0$, $e_n=\exp[-4/(\phi_1-\phi_0)^2]$. These initial conditions are under geostrophic balance, but with high potential for fluid instability. In order to trigger it, we add a perturbation to the height field:
\begin{equation}
    h'(\theta,\phi) = h_{\max}e^{-(\theta/\alpha)^2}e^{-[(\phi_2-\phi)/\beta]^2}\cos\phi,
\end{equation}
where $\phi_2=\pi/4$, $\alpha=1/3$, $\beta=1/15$, and $h_{\max}=120$ m. All schemes are tested on a g$_7$ refinement with a timestep of $50$ seconds under a RK4 timestepping scheme. In order to avoid the instability, we use a hyperviscosity coefficient of $5\times 10^{15}$ and $2\times 10^{15}$, for both A- and B-grid, respectively. These choices of coefficients are in agreement with \cite{TomitaSatoh2004}. We also found that smaller values of these coefficients of each scheme would lead to instability for the A-grid and the appearance of near grid scale oscillations in the B-grid.

\begin{figure*}[t]
    \centering
    \includegraphics[width=\linewidth]{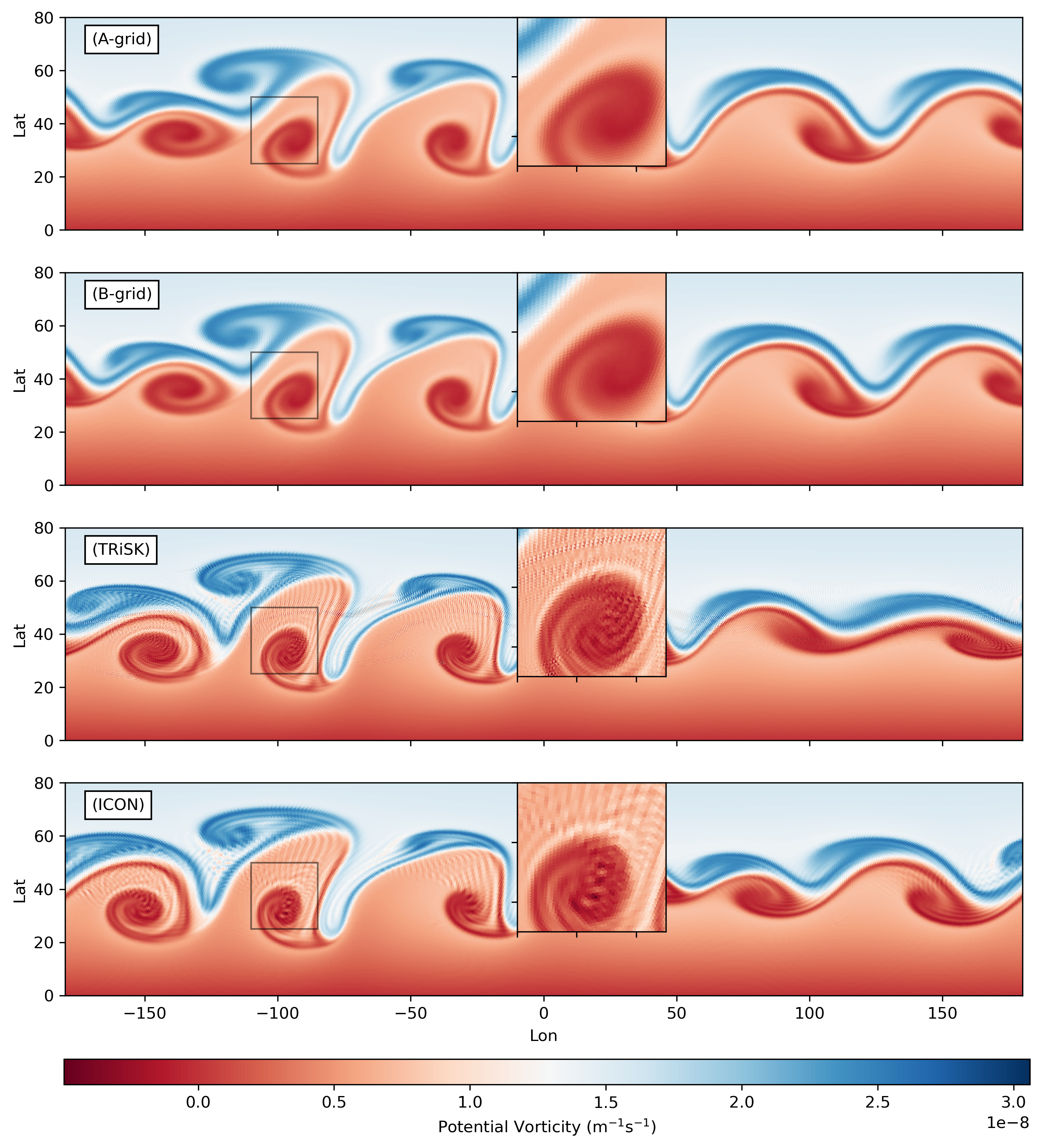}
    \caption{Potential Vorticity of all schemes on the 6th day of integration for the barotropic instability testcase with perturbation using a g$_7$ refinement grid and a respective biharmonic for A- and B-grid schemes, following Table \ref{tab:biharcoef}.}
    \label{fig:GalewskyPotVort}
\end{figure*}

The potential vorticity, on the sixth day of integration (Figure \ref{fig:GalewskyPotVort}), display the behaviour of the growth of the instability on all the evaluated schemes. Between these schemes, it is observed a clear difference in the representation of the smaller scale features of the instability. Both A-grid and B-grid schemes displays no small scale oscillations present within the vorticity field. Additionally, it is evident that both schemes display slightly coarser features in representing the state of the fields.

Similarly, in both C-grid schemes, we observe more small scale features in this system, helping could potentially aid in the growth of the instability even if no perturbation was added. However, it is evident that in these schemes, near-grid scale oscillations play a role in the physical solutions of the integration. Comparing both C-grid schemes, both schemes seem equally contaminated by numerical noise, however, the small scale oscillations in MPAS display a higher wavenumber than the ICON scheme. MPAS's noise in the vorticity was discussed and argued that the chequerboard noise of the vorticity is the main culprit in the manifestation of this contamination in our physical simulations \citep{Peixoto2016}. Likewise, we also know that the Perot's operator on the dual grid is capable of manifesting numerical noises on the solutions. Since ICON's divergence operator has the potential to remove small scale oscillations, but the scheme does manifest spurious waves, which was also observed in \cite{KornLinardakis2018}, therefore, the Perot's dual operator is potentially the main responsible for this manifestation.

Overall, all schemes suffer from the grid scale computational modes. There is, however, the stabilization term for both A- and B-grid schemes, such that the schemes remain stable throughout the integration. Despite both C-grid schemes remaining stable throughout the integration, the solutions are contaminated with noise, that will inevitably require a smoothing term, such as the biharmonic, in order to remove these high wavenumber waves. Additionally, It is observed that the waves from the A-grid to the C-grid schemes, an apparent increase in the effective resolution of the computation, agreeing  with the previous results in Section \ref{subsec:LNM}. Following this result, we analyse the kinetic spectrum of these schemes.

\subsubsection{Kinetic Energy Spectrum}

The global kinetic energy spectrum, is a useful tool in evaluating the energy cascade of the fluid. On different scales of the ocean's motion, we observe a power law of $k^{-3}$ for larger scales or $k^{-5/3}$ for smaller scales \citep{Wangetal2019}. For the 2D case, the former is related to the turbulence of the flow, whereas the latter is related to the reverse energy cascade turbulence. These spectral fluxes provide useful insight into the performance of the models in transferring energy motion between different scales.

Therefore, we define the Kinetic Energy Spectrum as follows: 
\begin{equation}
    (E_K)_n = \frac{a^2}{4n(n+1)}\left[\vert\zeta^0_n\vert^2 + \vert\delta^0_n\vert^2 + 2\sum_{m=1}^M\left(\vert\zeta^m_n\vert^2 + \vert\delta^m_n\vert^2\right)\right],
\end{equation}
where $\zeta^m_n$, $\delta^m_n$ are the spectral coefficient of the vorticity and divergence. These coefficients are defined as:
\begin{equation}
    \psi^m_n = \int_{-1}^1\frac{1}{2\pi}\mathcal{F}(\psi(\phi,\theta),\phi)\overline{P^m_n}(\theta)d\theta,
\end{equation}
where $\psi$ is the variable to be transformed, $\mathcal{F}(\psi(\phi,\theta),\phi)$ is the Fourier Transform on this variable, and $\overline{P^m_n}(\theta)$ is the normalized associate Legendre polynomial. To evaluate these equations, we use the nearest neighbour to interpolate the original unstructured grid into a quadrilateral grid of $10$ km resolution on the equator with the nearest neighbour method.

\begin{figure}[h]
    \centering
    \includegraphics[width=.7\linewidth]{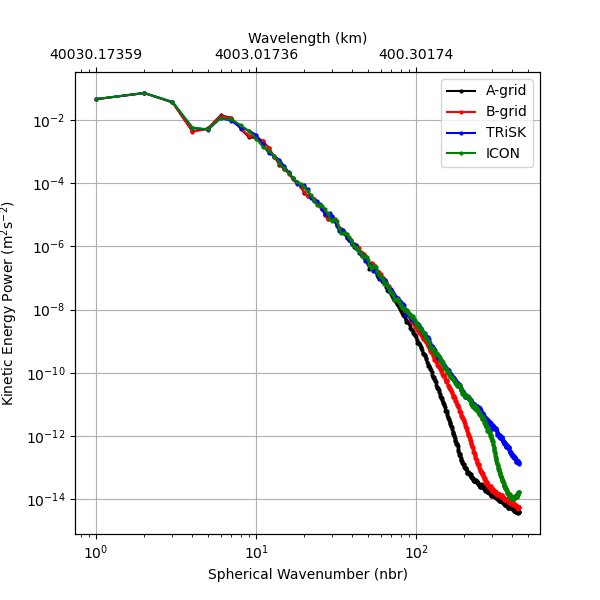}
    \caption{Kinetic energy spectra for the Barotropic instability testcase for all schemes as in Figure \ref{fig:GalewskyPotVort}.}
    \label{fig:KeSpecGalewsky}
\end{figure}

The energy spectrum of the schemes is shown on Figure \ref{fig:KeSpecGalewsky}. From the testcase, a small decrease of the spectrum from the wavenumber 1 to 4, and subsequently an increase, reaching a maximum at the wavenumber 6. Afterwards there is a constant decrease of the spectrum with a slope near $k^{-3}$ for all grids. At approximately wavenumber 80, the A-grid scheme has a considerable loss of its power, decreasing more rapidly. Similarly, at wavenumber 90 the B-grid scheme also displays this rapidly loss of energy. With slight higher wavenumber, both A- and B-grid slows its slope until the last evaluated wavenumber.

Comparably, both C-grid schemes extend the physical slope of $k^{-3}$ up to the wavenumber 300. At this wavenumber, ICON display a similar loss of kinetic energy, whereas MPAS maintain a similar slope up to the end of the evaluated wavenumbers. We again remark that our approach for ICON-O was to perform the mass lumping approach, which may have some impact on the effective resolution.

In summary, we have shown that for smaller wavenumbers there is a good agreement between the models. Additionally, we also have shown that even for the nonlinear time integration of the shallow water system of equations, the schemes behave similar to the linear normal mode analysis, with A-grid having the coarsest effective resolution, and MPAS, on the other extreme, having the highest effective resolution. Additionally, the presence of a slow-down of the loss of the power or even an increase of the spectrum on the highest wavenumbers is likely related to the impact of the interpolation to cause this increase, as it was previously reported in other works \citep{Wangetal2019, Ripodasetal2009, Jurickeetal2023}.

\FloatBarrier

\subsection{Models Stability}
\label{sec:HollsInst}

Our previous results were able to show elementary characteristics of each of the shallow water schemes. Some of our results required the inclusion of a stabilizing term for both A- and B- grid schemes, in order to remove damaging numerical oscillations that participated in the dynamics. Although the same term was not used in the C-grid scheme in our simulations, it is desired to include some sort of filtering, as the simulations may contain numerical waves that could either damage the solution or cause a potential \textit{blow up} of the model.

One particular cause of numerical dispersion is associated with 3D energy-enstrophy conserving models, regardless of the staggering used. The imbalance between the Coriolis and kinetic energy term generates numerical noise, causing near grid-scale oscillations and decreasing the kinetic energy of jets \citep{Hollingsworth1983}. This instability, known as Hollingsworth Instability, also manifests as a destabilized inertia-gravity wave, leading to a blow up of the solution depending on the models' resolution and distortion of the mesh \citep{Belletal2017, Peixotoetal2018}. Recent ocean models, such as NEMO's model, have shown susceptibility to these oscillations, producing spurious energy transfer to the internal gravity-waves and dissipation, resulting in corruption of mesoscale currents and submesoscale structures \citep{Ducoussoetal2017}.

Although this instability is 3D in nature, it is possible to mimic it, by considering the ocean model as a layered model, where the vertical flow is associated with one of the thin layers of the ocean \citep{Belletal2017}. This can be done by assuming the ocean model is hydrostatic and under a Bousinesq approximation (assumptions made by all ocean models evaluated in this work). In that case, one of the layers, henceforth equivalent depth $H$, if unstable, will display a strong noise on the horizontal velocity, and, thus, can be analysed with the shallow water equations.

\subsubsection{2D stability Analysis}

In order to examine the instability, we analyse the models under a nonlinear geostrophic testcase, similar to TC1. In this testcase, however, we consider the bathymetry as driving the geostrophic balance. The mass height field will be constant and small to mimic the equivalent depth of the internal modes of the 3D model, as done by \cite{Belletal2017}, and \cite{Peixotoetal2018}. Furthermore, we apply a linear analysis using the power method \citep{Peixotoetal2018}:
\begin{equation}
    \mathbf{x}^{(k+1)} = \alpha_{k+1}\mathbf{r}^{(k+1)}+\overline{\mathbf{x}},
\end{equation}
where $\alpha^{(k+1)}=\epsilon/\vert\mathbf{r}^{(k+1)}\vert$, $\epsilon=10^{-5}$ is a small constant, $\overline{\mathbf{x}}$ is the model state under geostrophic balance, $\mathbf{r}^{(k+1)} = \mathbf{x}^* - \overline{\mathbf{x}}$ is the perturbation, $\mathbf{x}^* = \mathbf{G}(x^{k})+\mathbf{F}$, $\mathbf{G}(x^{k})$ is the model evolution operator, and $\mathbf{F}=\overline{\mathbf{x}} - \mathbf{G}(\overline{\mathbf{x}})$ is a constant forcing. The methods converge, when $\alpha^k\to^k \alpha$ is found for large enough $k$. The eigenvalue is then obtained as $\lambda=1/\alpha$. From there we can compute the E-folding timescale from the growth rate  $\nu = \log\lambda/\Delta t$, where $\Delta t$ is the timestep. We will use, a timestep of $200$ seconds.

\begin{figure}[h]
    \centering
    \includegraphics[width=.7\linewidth]{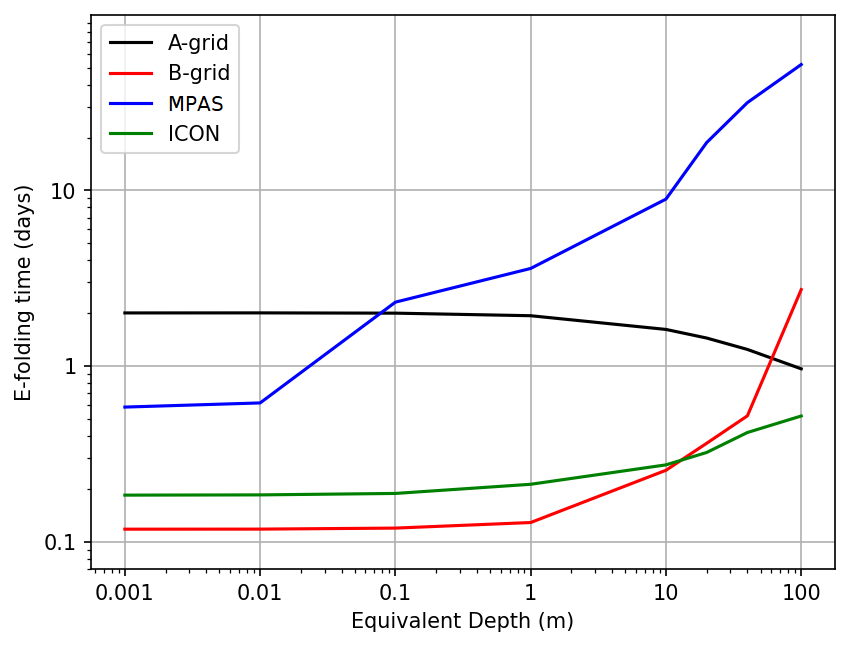}
    \caption{E-folding time for the different evaluated schemes, considering a time-step of 200 s in a geostrophic test case where the balanced state is given by the bathymetry, while the height is given by the equivalent depth and constant.}
    \label{fig:HollsInst}
\end{figure}

Ranging from an equivalent depth from $10^{-3}$ to $100$ m we observe a substantial difference between the stability of the evaluated schemes (Figure \ref{fig:HollsInst}). B-grid and ICON show similar e-folding time at around 0.1 and 0.2 days from the shallowest depth up to $1$ m. Larger thickness display a stabilization of both schemes. B-grid, in this case, display a faster stabilization than ICON, whose e-folding time remain below $1$ day for the $200$ m, whilst B-grid show over $2$ days e-folding time for the same thickness.

The similarities of both schemes for lower equivalent depths is potentially due to the use of triangular cells on some of their operators. However, the difference between the schemes for larger depths is likely associated with the error created by the reconstruction of the velocity vector field for both Coriolis and Kinetic energy terms in ICON, amplifying the imbalance of the discretization. Additionally, in different grids, ICON is found to be more stable \citep{KornLinardakis2018}, implying that our choice of grid might be a source of a higher instability.

On the other hand, both MPAS and A-grid display overall a more stable scheme. MPAS displayed a $0.6$ day e-folding time for the shallowest depths, but showed an increase, reaching around $40$ days. Similarly, A-grid displays an even larger stability of around $0.2$ day for the shallowest depth. However, contrary to the other schemes, the stability of the A-grid decrease with the increase of the equivalent depth. A-grid's stability loss with depth might be potentially due to different causes of instability being dominant for the equivalent depths, i.e. for shallower depths, the cause of the instability is likely the Hollingsworth Instability, while for deeper depths, the instability is caused by the excitation of spurious pressure modes.

\subsubsection{Biharmonic}

In order to evaluate the biharmonic effect on the stability of the models, we perform the same analysis for different viscosity coefficients, using an equivalent depth of $1$ metre, and a timestep of $200$ seconds. For A- and B-grid schemes, we use \eqref{eq:diffA} and \eqref{eq:diffB}, respectively. On C-grid, we use the formulation:
\begin{align*}
    \Delta\mathbf{u} = \nabla\nabla\cdot\mathbf{u}-\nabla\times\nabla\times\mathbf{u} \approx \textbf{grad }\textbf{div }u - \textbf{grad }^T\textbf{vort }u,
\end{align*}
where $\textbf{grad }^T$ is the transpose gradient operator defined on the dual grid.

\begin{figure}[h]
    \centering
    \includegraphics[width=.7\linewidth]{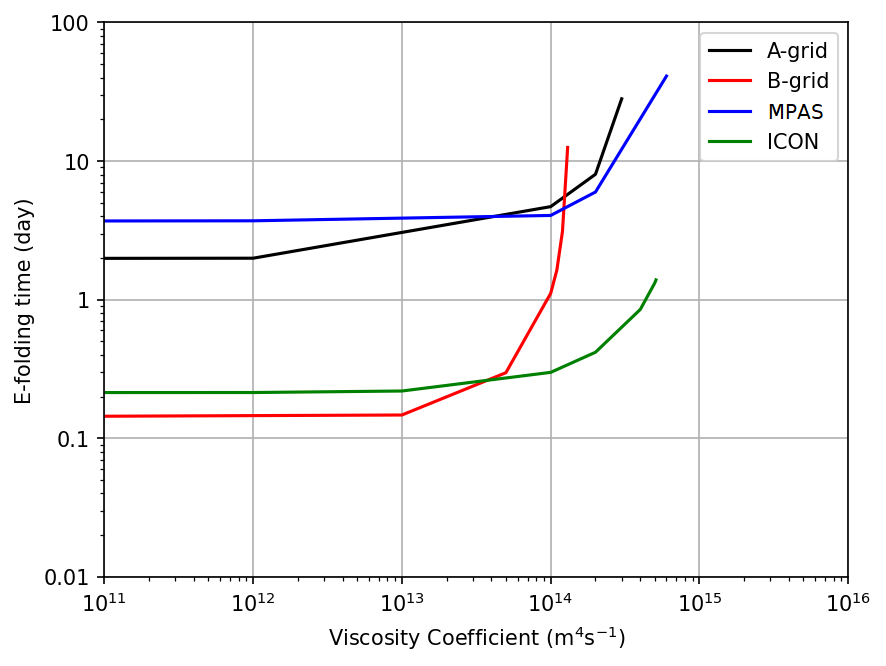}
    \caption{E-folding time by viscosity coefficient for each scheme, using a g$_6$ grid refinement with a timestep of 200 s and a 1 m equivalent depth.}
    \label{fig:HollsBiharm}
\end{figure}

Our analysis, shown on Figure \ref{fig:HollsBiharm}, indicates that all schemes were found to be stable for a viscosity coefficient no more than 10$^{15}$ m$^4$s$^{-1}$. Individually, B-grid and ICON does not display difference in stability for a coefficient up to 10$^{13}$ m$^4$s$^{-1}$. However, increasing the coefficient, shows that the B-grid has, not only a faster stabilization than ICON, but has the fastest of all evaluated schemes, reaching an e-folding time of over 10 days for a coefficient of $1\times 10^{14}$ m$^4$s$^{-1}$. ICON, in contrast, shows the slowest stabilization, reaching an e-folding time of 1.1 days for a coefficient of $4\times 10^{14}$ m$^4$s$^{-1}$.

Similarly, both A-grid and MPAS schemes display an unchanged e-folding time of up to $10^{13}$ m$^4$s$^{-1}$ and $10^{14}$ m$^4$s$^{-1}$, respectively. Additionally, A-grid is shown to stabilize faster than MPAS, reaching an e-folding time of over $20$ days for a coefficient of $3\times10^{14}$ m$^4$s$^{-1}$, while MPAS reaches $10$ days for the same coefficient.

Overall, we see that despite B-grid showing a lower stability than all schemes, it has the potential to faster achieve stability. Conversely, although ICON obtains a similar stability as the B-grid, it requires a more intense coefficient, in order to stabilize the scheme. The similar behaviour happens with A-grid and MPAS, with MPAS requiring a more intense coefficient for stabilization. This implies that this difficulty is on the C-grid discretization itself, and it is likely associated with either the vector reconstruction of the Coriolis term or the Kinetic Energy discretization.

\section{ICON-O Model}

Given the importance of the biharmonic term in order to stabilize the scheme or, at least, remove spurious computational waves in the system, we, then, aim to bridge the gap between the shallow water model and ICON's operational model. We will first acknowledge that our analysis in this section will be limited to ICON-O, and will not give light to other models mentioned in this work. However, providing results with ICON-O will be an important step towards understanding the effects of numerical oscillations on research/operational models. Additionally, our simulations presented in this section were not fine-tuned, i.e. the physical parameters and coefficients were not thoroughly calibrated, and, therefore, these simulations may not necessarily represent reality accurately. However, our discussions in this section will be focused on the analysis of the differences between simulations with and without the biharmonic filter, so the lack of calibration will not impact the overall analyses of the results.

The Ocean General Circulation Model ICON-O, developed at the Max-Planck Institute for Meteorology, is the oceanic component of the ICON Earth System Model. It uses horizontal discretization described in the earlier sections. Vertically, it extends the triangular cells into prisms, for the use of its z coordinate levels. Additionally, In its 3D formulation, ICON-O uses the hydrostatic and Bousinesq approximations to solve its state vector $\{u,h,T,S\}$, where $T$ and $S$ are temperature and salinity, respectively. These tracers are also imbued with dissipative and subgrid-scale operators, such as isoneutral diffusion and the mesoscale eddy advection Gent-Mcwilliams \cite{Korn2018}. The full 3D spatial discretization will be omitted in this section, but the reader can refer to equation (32) of \cite{Korn2017}.

For its time integration, ICON-O is discretized using an Adams-Bashforth 2-step predictor-corrector scheme (equation 33, 34, and 35 of \citep{Korn2017}). This timestepping scheme does not conserve neither energy nor enstrophy, and it provides an inherent diffusion \citep{KornLinardakis2018}.

\begin{figure}[h]
    \centering
    \includegraphics[width=.8\linewidth]{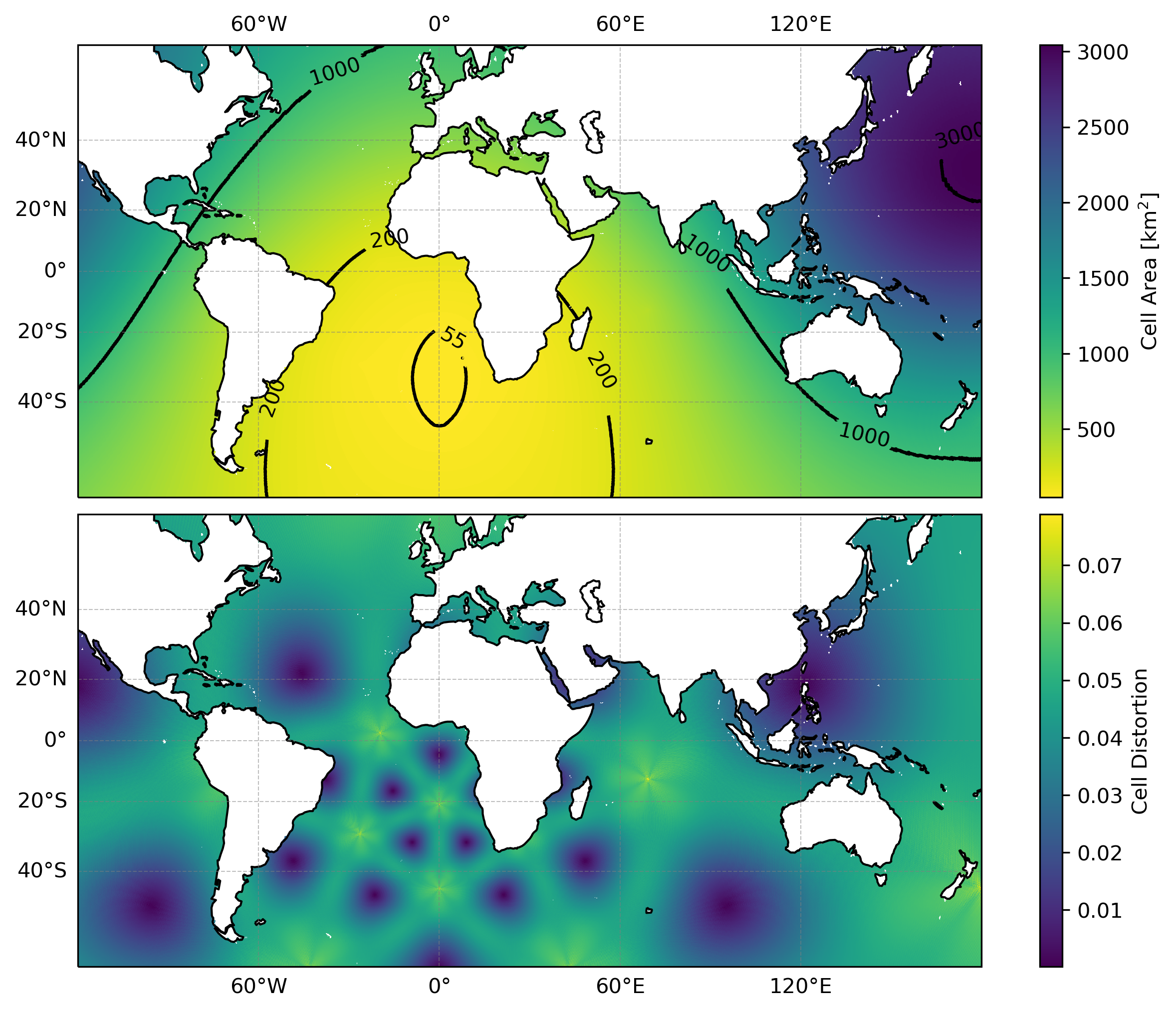}
    \caption{The upper panel is the cell area of the spherical grid used in the simulations. The lower panel is the respective cell distortion of the mesh.}
    \label{fig:areaDist}
\end{figure}

Our 3D simulations were performed using a Spring Dynamics optimized grid with a radial local refinement with the finest resolution, around 14 Km edge length, located near South Africa, and the coarsest resolution, around 80 Km edge length, on the antipode of the earth, i.e. North Pacific (Figure \ref{fig:areaDist} upper panel). These locally refined mesh created enumerated distortion spots around the refined region (Figure \ref{fig:areaDist} lower panel).

The model was initialized under rest with $128$ layers with climatological temperature and salinity from the Polar Science Center Hydrographic Climatology \citep{Steeleetal2001} and was forced with the German-OMIP climatological forcing, which is derived from the ECMWF reanalysis 15 years dataset. This climatological forcing is daily with a resolution of 1 degree. An initial thirty years spin up was performed under these conditions utilizing a biharmonic coefficient of $2\times 10^{-1} A_e^{3/2}$, where $A_e=\vert e \vert\vert \hat{e} \vert/2$. In addition, we added a Turbulent Kinetic Energy (TKE) closure scheme, for the vertical diffusivity of traces and velocities.

Following the spin up, we, subsequently, ran 2 simulations by $10$ years each. One simulated with the same parameters as the spin up, which we will coin as our reference simulation. The other was simulated without the aforementioned biharmonic filter, which we will coin as NB run.

\begin{figure*}
    \centering
    \includegraphics[width=\linewidth]{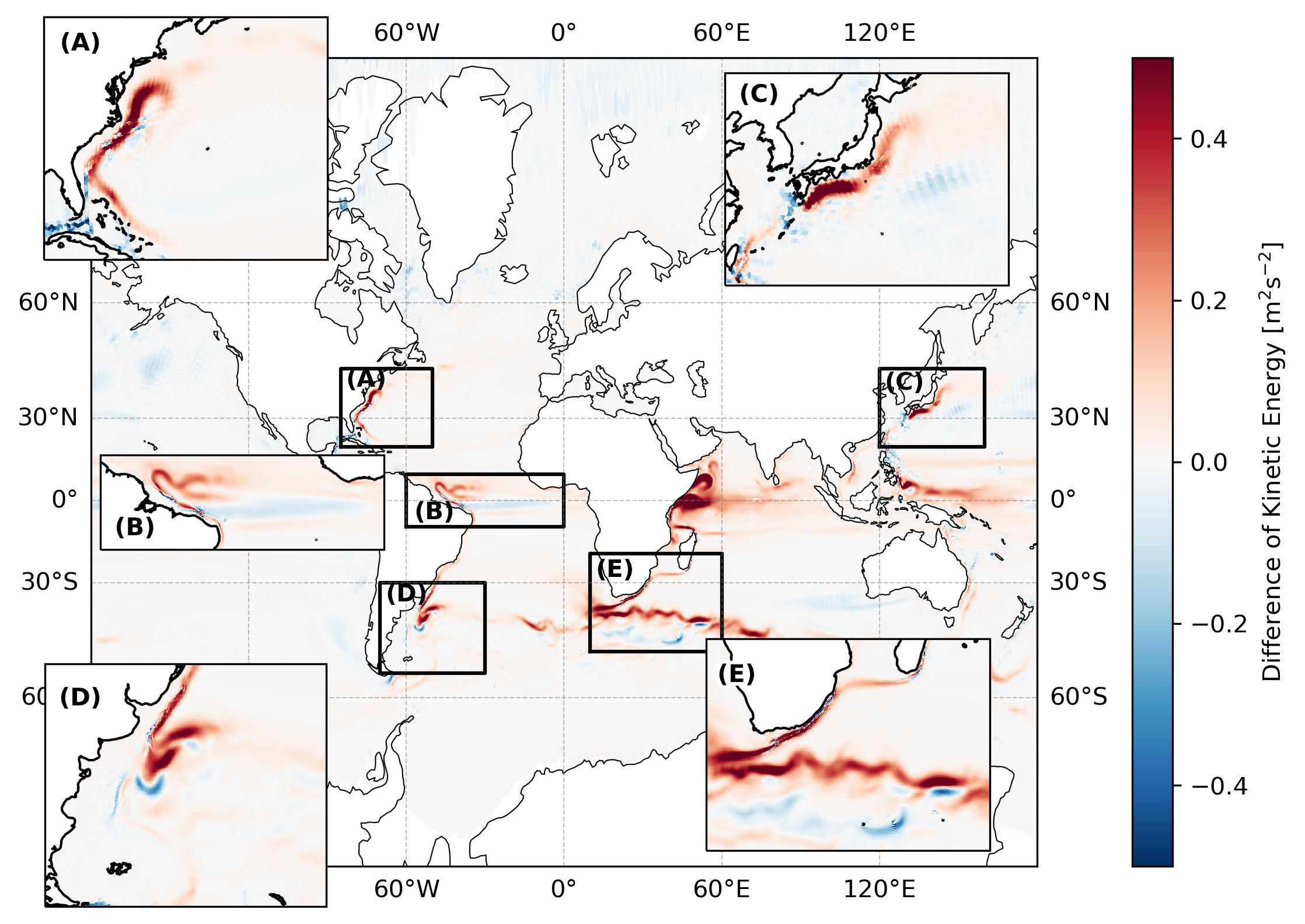}
    \caption{Kinetic Energy difference between a reference simulation and simulation without the use of biharmonic, i.e. $E_k^{(\text{ref})}-E_k^{(\text{no bih})}$.}
    \label{fig:kineticenergy}
\end{figure*}

The simulation without the filter show a clear decrease in the strength of the currents on the ocean system (Figure \ref{fig:kineticenergy}, e.g. Gulf Stream (A), North Equatorial (B), Kuroshio (C), Malvinas currents (D), and Agulhas (E)). Other regions were found to slightly increase in kinetic energy, in particular, the neighbourhood around the Agulhas Current, near the Antarctic Circumpolar Current, the Equatorial Currents of the Atlantic Ocean and both Northern and Southern of the Pacific Ocean, and the Brazil-Malvinas Confluence. The integrated kinetic energy averaged over these years show that surface kinetic energy loss of around 4.7 $\times10^{13}$ km$^2$m$^2$s$^{-2}$ of its 20 $\times10^{13}$ km$^2$m$^2$s$^{-2}$. Additionally, it is observed, in particular on regions of coarser resolution, such as the Kuroshio Current and Gulf Stream, the presence of a numerical oscillation emanating from the main currents.

\begin{figure}
    \centering
    \includegraphics[width=.6\linewidth]{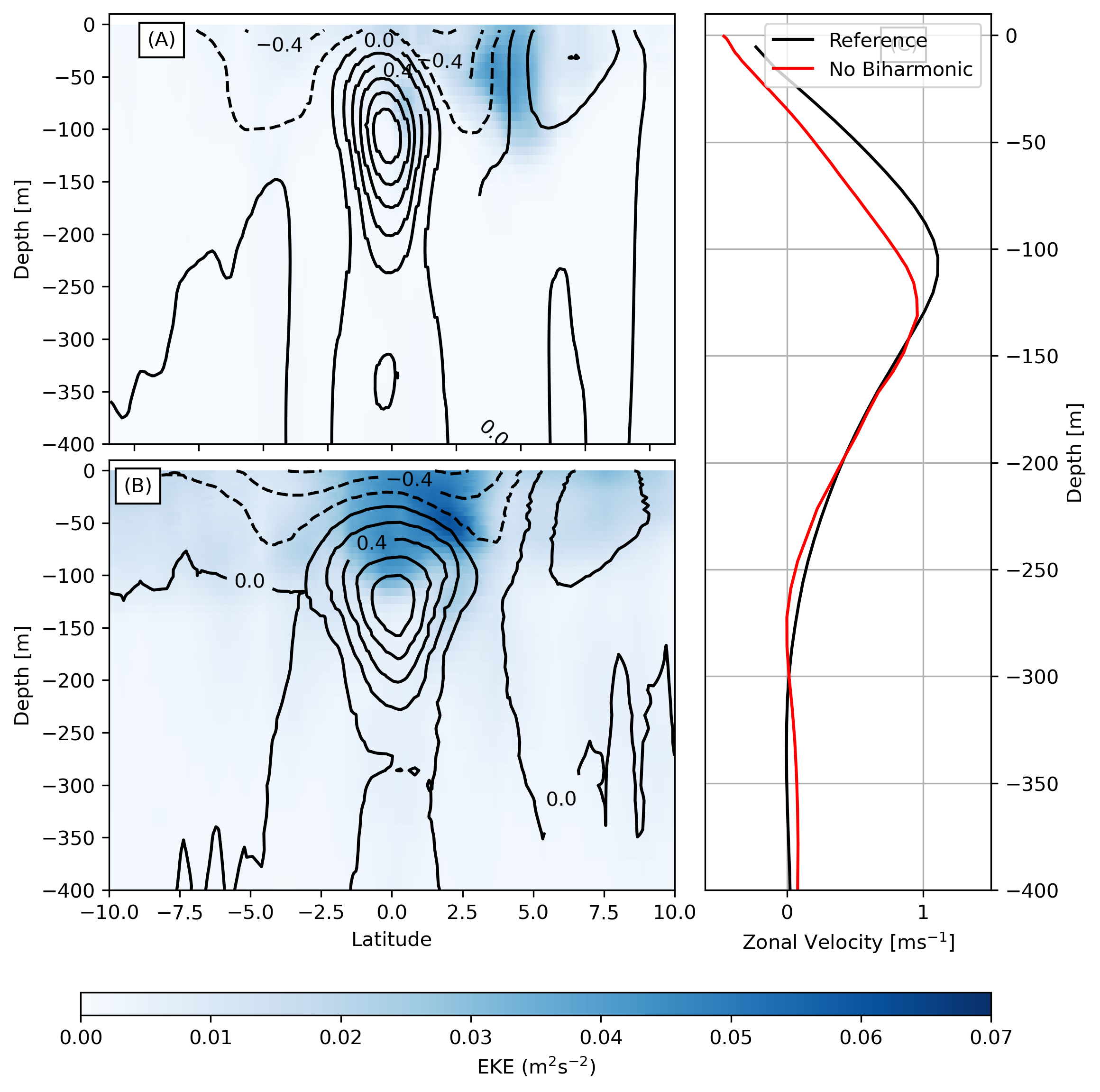}
    \caption{Cross-section of the 130$^\circ$W longitude of the reference (A) and the without biharmonic (B) simulation and a vertical profile of the zonal velocity of both simulation over the 0$^\circ$ Latitude (C).}
    \label{fig:crosssec130W}
\end{figure}

At the equatorial pacific currents,  in our experiments, we observe that the NB simulation show a wider jet with a weaker and deeper core intensity (Figure \ref{fig:crosssec130W}). Moreover, the NB simulation show that the northern and souther branches of the Equatorial Current decrease in their intensity, and a flow intensity up from the EUC, which likely occurs due to the deepening  of the EUC. In relation to the turbulent energy, the NB simulation shows an increase of EKE at the interface between the slow westward surface flow and the EUC, while decreasing its EKE at the northernmost edge of the North Equatorial Current. \cite{Ducoussoetal2017} in their work on NEMO also observed a deformation of the equatorial undercurrent, however, in their experiments, the current was shown to narrow vertically, and they overall detected a decrease in the EKE field. According to the authors, this effects occur because the region is highly dependent of the baroclinic instability. According to the authors, this system of currents is highly subject to baroclinic instabilities, generating waves and eddies which are the main contributors of the current. The decrease in intensity of the currents could be explained to the decrease in baroclinic instabilities. Similarly, the increase in EKE detected in NB are potentially explained by either a shear between both EUC and the newly generated surface flow and/or by a spurious mixing caused by the emission of numerical oscillation which draws energy from the currents to provide mixing between the both layers.

\begin{figure}
    \centering
    \includegraphics[width=.6\linewidth]{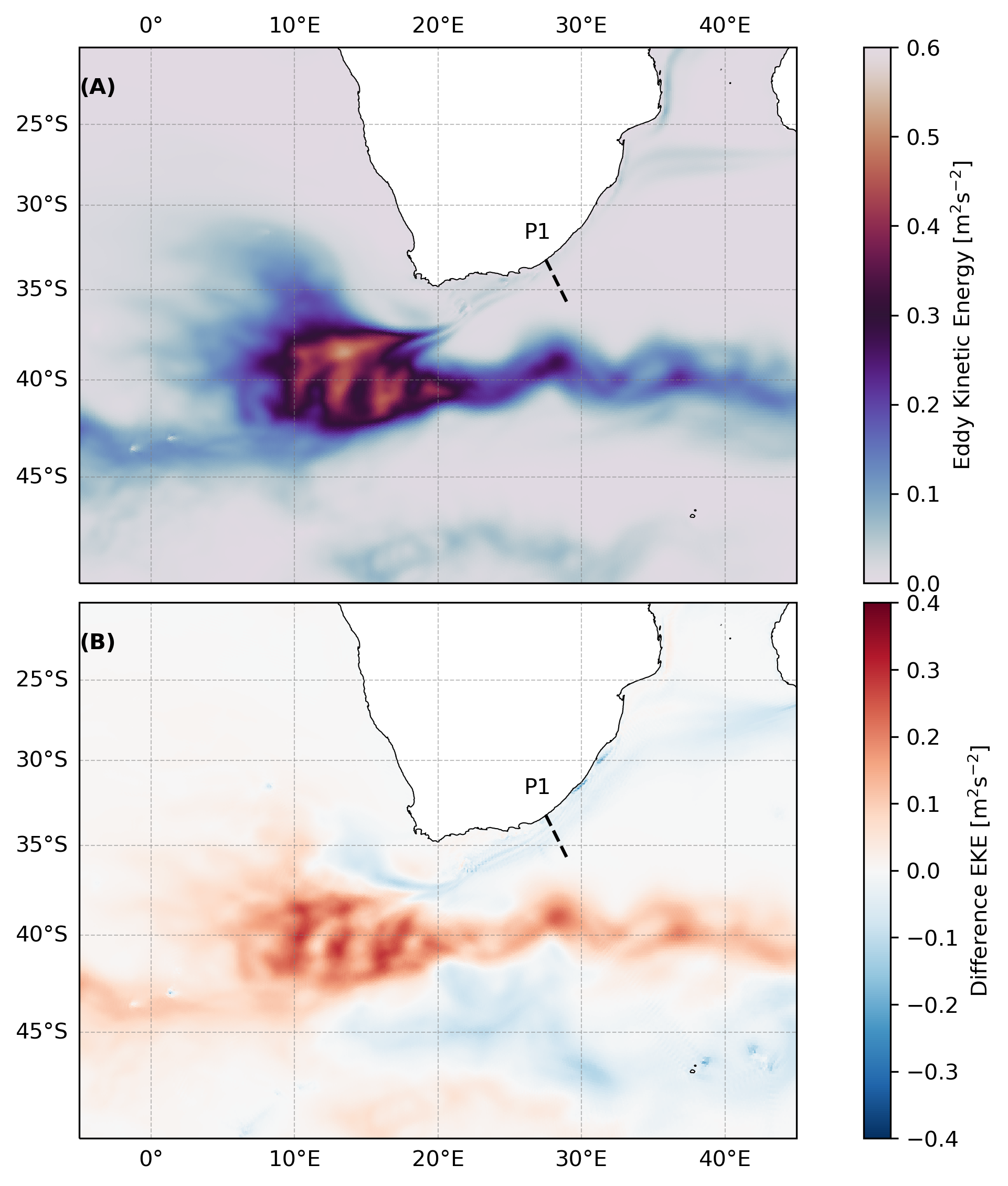}
    \caption{Eddy Kinetic Energy (A) and difference between simulations of EKE (B) of the Agulhas Current System.}
    \label{fig:ekediff}
\end{figure}

A similar EKE effect is detected on other oceanic regions. Most notably at the Agulhas Current Retroflection, where it meets with the colder water of the South Atlantic Current and Antarctic Circumpolar Current (Figure \ref{fig:ekediff}). The retroflection region EKE is known to be modulated by the baroclinic instability of the Agulhas current \citep{Zhuetal2018}. 

\begin{figure}
    \centering
    \includegraphics[width=.6\linewidth]{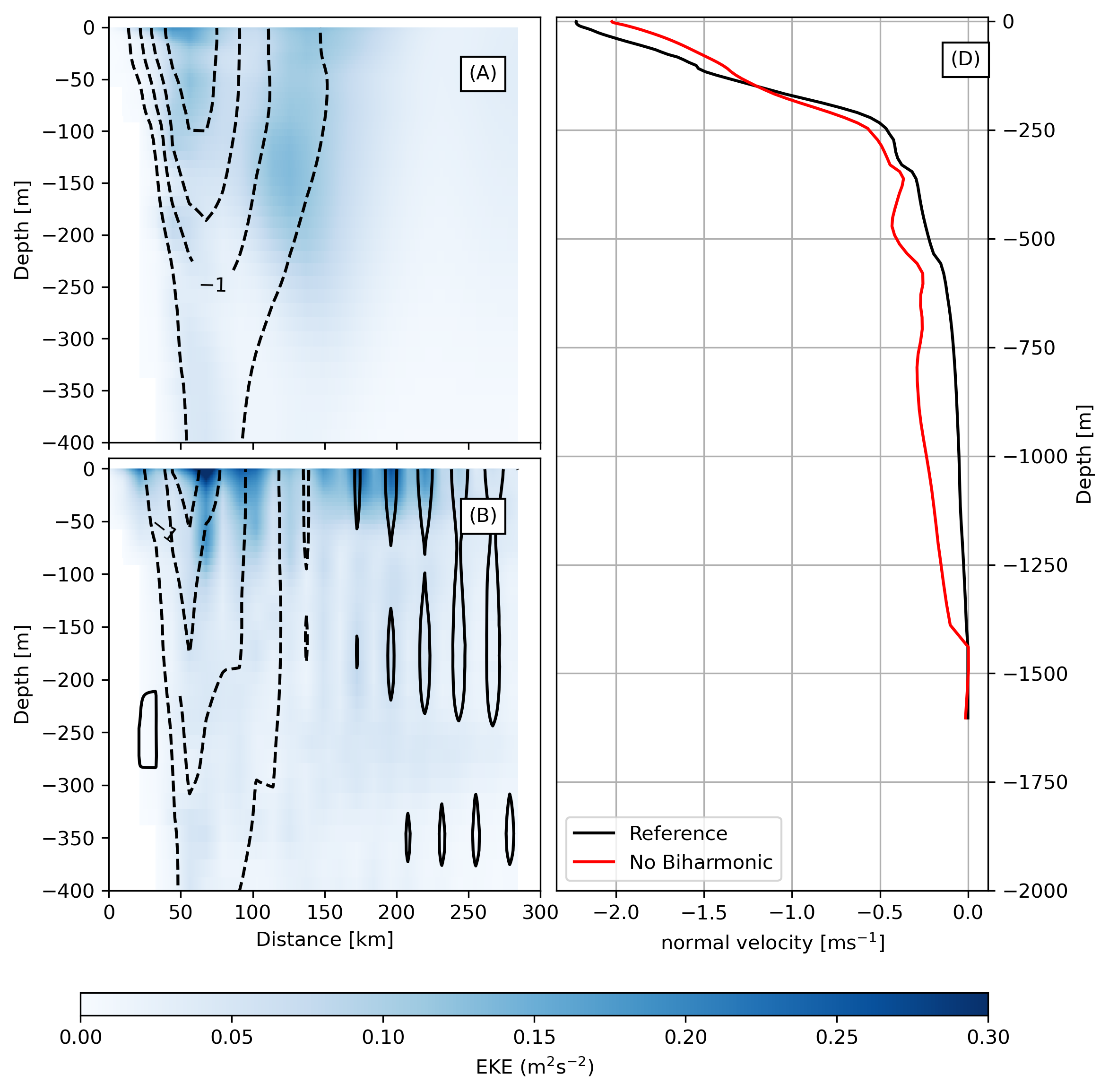}
    \caption{P1 Cross-section between the Observational data (A), Reference simulation (B), and No Biharmonic Simulation (C), and the vertical profile of the normal velocity in the 42 km distance (D).}
    \label{fig:csecag}
\end{figure}

Additionally, at the Agulhas Current itself, where there is less intensity in the EKE, the NB simulation shows a slight increase of this field. Observing the cross-section P1, we note a clear decrease in intensity of jets core (Figure \ref{fig:csecag}.C) at the surface, while a weak normal flow is generated at the higher depths. Additionally, it is observed that the NB simulation generate small scale flow spanning near the whole water column, manifesting from the Agulhas Current and propagating tangent of the cross-section (Figure \ref{fig:csecag}.B). It is likely that these oscillations are responsible for the increase in EKE of the field at the core of the current and, consequently, the decrease of the intensity of jet, which may overall impact on the Agulhas Current Retroflection intensity.

\FloatBarrier

\section{Conclusions}

In this work, we provided a thorough comparison analysis between different shallow water staggering schemes used in unstructured ocean models and their capability in maintaining a stable integration. Alongside, we also investigated ICON's susceptibility to such numerical instabilities in realistic 3D settings.

The shallow water analyses have shown that all models haves advantages and disadvantages. The NICAM horizontal discretization, from \cite{Tomitaetal2001}, is simple to discretize, due to its collocated approach, provides accurate representation of the operators, and presents reasonably stable integrations for complex experiments, for chosen grid optimizations, such as the SCVT. However, similar to the traditional discretization of A-grids on regular grids \citep{ArakawaLamb1977, Randall1994}, it displays a low effective resolution, difficulty in maintaining the geostrophic balance, and it is susceptible to the manifestation of numerical oscillations caused by the grid discretization.

Similarly, the FeSOM 2.0 horizontal discretization, from \cite{Danilovetal2017}, also provides a quite simple discretization, accurate approximations of the operators, and a higher effective resolution compared to the A-grid. However, it also has a low effective resolution, and it displays some difficulty in maintaining the geostrophic balance. Additionally, despite not suffering from pressure modes, the B-grid scheme is found to be the least stable scheme, but as shown here and discussed by \cite{Danilov2013}, It can be easily fixed by a low coefficient of biharmonic.

Finally, both C-grid schemes, MPAS-O, from \citep{Skamarocketal2012}, and ICON-O, from \cite{Korn2017}, have the most complex discretizations between the evaluated schemes. Some operators do not accurately approximate the operators of the Shallow Water system. The difficulty for MPAS-O to show convergence in the error was also discussed by \cite{Peixoto2016}. Similarly, ICON-O also displays some difficulty in converging some of the operators of the shallow water equations. The lack of convergence of the divergence operator, for example, was also shown in \cite{KornLinardakis2018} for their defined Rossby Grid. Therefore, for both schemes, it is argued that the issue lies in the use of the grid. Therefore, a proper choice of grid optimization should also be taken into consideration when using or using these schemes. Additionally, the difference in apparent effective resolution is observed for both grids, with MPAS-O having a higher resolution. This may be explained by the use of the grid optimization, the mass lumping approach or the Perot operator in ICON-O. Finally, a dissimilarity between both schemes is seen in their stability. MPAS is shown to have a high stability, as it was discussed in \citep{Peixotoetal2018}, but ICON, similar to the B-grid, is shown to have a low stability and requires a larger viscosity than B-grid to stabilize the scheme. The grid use and the mass lumping may again be responsible for this difference. Despite this, a comparison between the use of difference computation of each operation is welcome to analyse how ICON-O's stability is impact, e.g. a comparison between the naive and Perot's computation of the divergence, kinetic energy, and perpendicular velocity.

Remarkably, in the 3D ICON-O simulation using a grid with Spring Dynamics optimization, the model was found to be stable throughout the simulated years, despite the lack of biharmonic filter. However, near grid oscillations were apparent in the grid and a contribution of these oscillations of the dynamics of the model was apparent. As it was also diagnosed by \cite{Ducoussoetal2017} for the NEMO model, these oscillations seemed to give rise to spurious mixing of the system and also decreases the energy of the ocean's currents. Regions where its strength is derived from baroclinic instability seems more affected by these small scale oscillations. Yet, it is clear the need for further research in this topic. Though the model is stable, it can be affected by these oscillations if the coefficient is not properly adjusted. Moreover, an excess of the viscosity may also decrease the effective resolution of the model, which also is not ideal.

In conclusion, we stress that further research is necessary in order to shed more light into these schemes. We note that all schemes under the shallow water tests have shown to be robust and provide reliable results for their respective purpose. However, testing these schemes under different grids or with more realistic settings might provide greater insights into the performance of the models. Additionally, it seems evident that despite a model being stable without filters, the numerical oscillations in the model may interact with the physical waves, leading to errors or to misinterpretation of the results. It is, therefore, crucial for further investigation on this topic in order to properly make use of filters to avoid these oscillations, but also minimize the damping of physical waves. 

\section{Acknowledgements}
We are grateful to the financial support given by the Brazilian Coordination for the Improvement of Higher Education Personnel (CAPES) PRINT project - Call no. 41/2017, Grant 88887.694523/2022-00, the São Paulo Research Foundation (FAPESP) Grant 2021/06176-0, and the Brazilian National Council for Scientific and Technological Development (CNPq), Grants 140455/2019-1 and 303436/2022-0.
\FloatBarrier


 \bibliographystyle{elsarticle-harv} 
 \bibliography{elsarticle-template-num}





\end{document}